\documentclass[aps,prb,reprint,superscriptaddress,showpacs,longbibliography,floatfix,footnoteinbib]{revtex4-2}
\usepackage{amsmath,amssymb,graphicx,units}
\usepackage[plainpages=false,pdfpagelabels,colorlinks=true,linkcolor=red,urlcolor=blue,citecolor=blue,pdftitle={Title},pdfauthor={},pdfdisplaydoctitle=true,pdfduplex=DuplexFlipLongEdge]{hyperref}
\usepackage{multirow}
\usepackage{setspace}
\usepackage{bm}
\usepackage{lipsum} 

\begin{document}

\title{On the criticality of the configuration-space statistical geometry}

\author{Yu-Jing Liu}
\affiliation{Lanzhou Center for Theoretical Physics, Key Laboratory of Quantum Theory and Applications of MoE, Key Laboratory of Theoretical Physics of Gansu Province, and Gansu Provincial Research Center for Basic Disciplines of Quantum Physics, Lanzhou University, Lanzhou, Gansu 730000, China}

\author{Wen-Yu Su}
\affiliation{Lanzhou Center for Theoretical Physics, Key Laboratory of Quantum Theory and Applications of MoE, Key Laboratory of Theoretical Physics of Gansu Province, and Gansu Provincial Research Center for Basic Disciplines of Quantum Physics, Lanzhou University, Lanzhou, Gansu 730000, China}

\author{Yong-Feng Yang}
\affiliation{Lanzhou Center for Theoretical Physics, Key Laboratory of Quantum Theory and Applications of MoE, Key Laboratory of Theoretical Physics of Gansu Province, and Gansu Provincial Research Center for Basic Disciplines of Quantum Physics, Lanzhou University, Lanzhou, Gansu 730000, China}

\author{Nvsen Ma}
\email{nvsenma@buaa.edu.cn}
\affiliation{School of Physics, Beihang University, Beijing 100191, China}
\affiliation{Lanzhou Center for Theoretical Physics, Key Laboratory of Quantum Theory and Applications of MoE, Key Laboratory of Theoretical Physics of Gansu Province, and Gansu Provincial Research Center for Basic Disciplines of Quantum Physics, Lanzhou University, Lanzhou, Gansu 730000, China}

\author{Chen Cheng}
\email{chengchen@lzu.edu.cn}
\affiliation{Lanzhou Center for Theoretical Physics, Key Laboratory of Quantum Theory and Applications of MoE, Key Laboratory of Theoretical Physics of Gansu Province, and Gansu Provincial Research Center for Basic Disciplines of Quantum Physics, Lanzhou University, Lanzhou, Gansu 730000, China}

\begin{abstract}
While phases and phase transitions are conventionally described by local order parameters in real space, we present a unified framework characterizing the phase transition through the geometry of configuration space defined by the statistics of pairwise distances $r_H$ between configurations. Focusing on the concrete example of Ising spins, we establish crucial analytical links between this geometry and fundamental real-space observables, i.e., the magnetization and two-point spin correlation functions. This link unveils the universal scaling law in the configuration space: the standard deviation of the normalized distances exhibits universal criticality as $\sqrt{\mathrm{Var}(r_H)}\sim L^{-2\beta/\nu}$, provided that the system possesses zero magnetization and satisfies $4\beta/\nu < d$. We validate this scaling with stochastic series expansion quantum Monte Carlo simulations of the transverse-field Ising model (TFIM). Furthermore, we propose configuration-space diagnostics that go beyond local real-space observables. First, the distribution probability $P(r_H)$ parameterized by the transverse field $h$ forms a one-dimensional manifold. Information-geometric analyses, particularly the Fisher information defined on this manifold, successfully pinpoint the TFIM phase transition, regardless of the measurement basis. Second, for the Su-Schrieffer–Heeger Heisenberg model, a parity index derived from $P(r_H)$ successfully characterizes the symmetry-protected topological phase and its transition. Our work establishes configuration space geometry as a novel perspective on quantum criticality, revealing how macroscopic universal phenomena are encoded within its global statistical features.

\end{abstract}
\maketitle

\section{Introduction}

Phase transitions, representing profound transformations between distinct states of matter under continuous variation of external parameters, constitute a cornerstone challenge in statistical physics and condensed matter physics, revealing universal principles governing collective behavior across various classical and quantum systems~\cite{cardy1996}. The Landau-Ginzburg-Wilson (LGW) paradigm~\cite{Landau1999, Sachdev_2011, Wilson1974} stands as the foundational theoretical framework for understanding continuous phase transitions. At its conceptual core lies the identification of an order parameter, a quantity manifesting the broken symmetry of the system, which underpins the entire description of the transition. Key elements of this paradigm include spontaneous symmetry breaking and the emergence of universal critical behavior, ultimately governed by the renormalization group (RG) theory~\cite{Wilson1971, Wilson1974}. Crucially, even in transitions beyond the strict applicability of the LGW universality predictions, such as the Berezinskii-Kosterlitz-Thouless (BKT) transition in the 2D $XY$ model~\cite{Kosterlitz1973, Kosterlitz1974}, the identification and analysis of a suitable order parameter remain indispensable for understanding the underlying physics, including the nature of symmetry breaking and the dynamics of topological excitations.

Although the order parameter provides profound insight into phase transitions, phases and transitions can be characterized through fundamentally distinct paradigms. On the one hand, significant classes of systems evade description via symmetry breaking and conventional order parameters: topological phase transitions are governed by changes in global topological invariants~\cite{Thouless1982, Kane2005}, many-body localization (MBL) is diagnosed through statistics of eigenlevels and eigenstates~\cite{Nandkishore2015, MBL_RMP2019, Logan2019, Sutradhar2022, Roy2024, Cheng2023}. On the other hand, machine learning (ML) offers a data-driven approach transcending predefined order parameters: these methods detect phase transitions directly from raw configuration data~\cite{Carrasquilla2017, vanNieuwenburg2017,ising3,santos1, Santos2021prxq, Giataganas_2022,arnoldprx}—even in regimes where order parameters are ill-defined, such as disordered systems~\cite{mbl_ml_2019,mbl_ml_2024, Fan2023} and topological phase transitions~\cite{Beach2018, Rodriguez-Nieva2019}. Critically, modern ML techniques shift the paradigm from real-space local observables to the geometry of high-dimensional configuration spaces, extracting universal signatures through latent representations.

In ML approaches, information about phases and phase transitions is extracted from various data mining techniques, which often act as ``black boxes'' and obscure the physical mechanisms behind their
predictions. To bridge this gap, one may confront the geometry of configuration spaces and unveil the secrets of phases and phase transitions hidden within it. In prior work~\cite{Su2024}, we demonstrated this directly: phase transitions across classical spin models with diverse universality classes were successfully characterized by quantifying the distribution of microscopic configurations and their pairwise Hamming distances. This approach bypasses complex ML pipelines, establishing configuration space geometry as a fundamental probe for universal criticality. However, these successes, derived from abundant numerical tests, demand deeper theoretical exploration: what is the physical interpretation of critical exponents extracted from configuration-space correlations and their physical connection to real-space order parameters? 

Motivated by the above question, as well as the successes of ML approaches~\cite{Carrasquilla2017, vanNieuwenburg2017, ising3, santos1, Santos2021prxq, Giataganas_2022, arnoldprx, mbl_ml_2019, mbl_ml_2024, Fan2023, Beach2018, Rodriguez-Nieva2019} and MBL characterization~\cite{Nandkishore2015, MBL_RMP2019, Sutradhar2022, Logan2019, Cheng2023, Roy2024} of phases and phase transitions beyond local order parameters, in this work, we propose a comprehensive framework for characterizing phase transitions in the Hilbert space, based on the geometry of the statistics of distances between configurations. We achieve this objective by following straightforward steps: In Sec.~\ref{sec:derivations}, after introducing basis definitions and notations, we establish an analytical relationship between the statistical moments of configuration-space distances and fundamental real-space observables, revealing that the former serves as an order parameter as long as the real-space correlation captures the phase transition. In particular, the standard deviation of the distance has a universal scaling law, provided the criterion $4\beta/\nu < d$ is satisfied. In Section~\ref{sec:numerics}, the analytical framework is verified using numerical quantum Monte Carlo (QMC) simulations of phase transitions in the transverse field Ising model (TFIM): the standard deviation of the distance captures the criticality in 1D and 2D TFIM, although the criterion is marginally violated in the 2D case; in  contrast, it fails when measured along the orthogonal direction of the symmetry breaking, where the real-space correlation does not capture the spontaneous symmetry breaking. In Sec.~\ref {sec:nonlocal_probes}, we propose and discuss configuration-space measurements that are beyond the real-space local observables: i) the Fisher information, viewed from the perspective of information geometry, which acts as a global probe of criticality, regardless of the choice of the measurement basis; ii) the parity index of the distance distribution, which is able to capture the symmetry-protected topological phase and phase transition. 

By shifting the perspective from traditional real-space observables to the geometry of high-dimensional configuration space, the present work demonstrates how the latter captures the criticality in various quantum spin systems. We hope that the present work inspires future investigations along this line.

\section{Distance in configuration space, low order moments and criticality}
\label{sec:derivations}

For a specific system described by a mixed or pure state, its underlying physics is determined by the corresponding partition function or wavefunction. In most cases, the information of phases and phase transitions can be extracted from the distribution of configurations $P(s)$, where $|s\rangle$ is a configuration in the chosen basis of the many-body Hilbert space. However, in practical problems involving many particles, the explicit distribution $P(s)$ can hardly be determined or even characterized, as the degrees of freedom scale exponentially with the lattice size. An alternative way is to consider the distribution of pairwise distances between configurations~\cite{Su2023, Su2024}, whose degrees of freedom are proportional to the size of the system. Our proposal relies on the fundamental assumption that the latter distribution contains the essential information on phases and phase transitions. 

In the following, without loss of generality, we take the system with Ising spins as an example to introduce the preliminaries of the present work, including basic definitions, notations, and derivations related to the criticality and statistical geometry of the configuration space. The derivation and conclusion can be easily extended to all lattice models with two-level local sites, such as spinless fermions or hard-core bosons.  

\subsection{Distances in configuration space}

Given two configurations $|s\rangle = |s_1, s_2, \dots, s_N\rangle$ and $|s^\prime\rangle = |s^\prime_1, s^\prime_2,  \dots, s^\prime_N\rangle$ sampled from the equilibrium ensemble of a lattice system with size $N$, the normalized pairwise distance in the Hilbert space is defined as 
\begin{align}
    \label{eq:distance}
    r_H(s, s^\prime) \equiv R_H(s, s^\prime)/N = \sum_{i=1}^N \frac{1 - s_i s^\prime_i}{2N}.
\end{align}
Here $s_i, s^\prime_i \in \{+1, -1\}$ are single-site eigenvalues in a chosen measurement basis, i.e., $\langle s_i\rangle=\langle \hat{\sigma}^\gamma_i \rangle$ in $\hat{\sigma}^\gamma$ basis with $\hat{\sigma}^\gamma$ ($\gamma=x,y,z$) is the Pauli matrix. The label $r_H$ ($R_H$) denotes the distances in the Hilbert space, differentiating from distances $r$ in real space. Since all configurations form a complete basis for the Hilbert space, we treat the space of configurations and the Hilbert space as equivalent; that is, these terms are used interchangeably throughout this paper.

The definition in Eq.~\eqref{eq:distance} is actually the widely used Hamming distance, which contains the equivalent information as the Euclidean distance in the case of lattices with two local levels~\footnote{The Hamming distance is actually the 1-norm distance between two configurations, while the Euclidean distance is the 2-norm, defined as $r_E(s,s') = \sqrt{\sum_{i=1}^N (s_i - s'_i)^2}$~\cite{Facco2017,Rodriguez-Nieva2019,santos1,Santos2021prxq}.  For systems with two-level local sites, the two distances contain equivalent information because the spin variables are binary.}.
This concept has been widely used in various ML approaches~\cite{Rodriguez-Nieva2019,santos1,Santos2021prxq} and MBL problems~\cite{Logan2019,Sutradhar2022,Guo2021,Yao2023,Roy2024}. More specifically, the distance distribution $P(r_H)$, as well as its low-order moments, i.e., the mean $\langle r_H \rangle$ and the standard deviation $\sqrt{\text{Var}(r_H)} = \sqrt{\langle r_H^2 \rangle - \langle r_H \rangle^2}$, have been numerically demonstrated to be closely related to the phase transition in previous studies~\cite{Su2023, Su2024, Yi2022}. In the rest of this section, we will unveil the underlying physics of numerical success by bridging the statistical measurements in configuration space with the conventional real-space observables. 


\subsection{Mean and variance of the configuration-space distance}

The expectation value $\langle r_H \rangle$ can be obtained by averaging all possible pairs of configurations drawn from the system's equilibrium ensemble. Adopting the linearity of expectation value, we have
\begin{align}
    \label{eq:rH_expr}
    \langle r_H \rangle = \left\langle \sum_{i=1}^N \frac{1 - s_i s'_i}{2N} \right\rangle = \sum_{i=1}^N \frac{1 - \langle s_i s'_i \rangle}{2N}.
\end{align}
Since the two configurations $|s\rangle$ and $|s^\prime\rangle$ are independently sampled from the same ensemble, one can factorize the expectation value of the product as
\begin{align}
    \langle s_i s'_i \rangle = \langle s_i \rangle \langle s'_i  \rangle = \langle \hat{\sigma}^\gamma_i \rangle^2.
\end{align}
Substituting this into Eq.~\eqref{eq:rH_expr} one gets the mean of distances as 
\begin{align}
    \langle r_H \rangle = \frac{1}{2} \left(1 - \frac{1}{N}\sum_i\langle \hat{\sigma}^\gamma_i \rangle^2 \right),
    \label{eq:mean_rH_0}
\end{align}
Assuming homogeneous $\langle \hat{\sigma}^\gamma_i \rangle$ in the real space, which is satisfied for most clean systems, the above equation simplifies to 
\begin{align}
    \langle r_H \rangle = \frac{1}{2}(1 - m_\gamma^2)
    \label{eq:mean_rH}
\end{align}
with $m_\gamma \equiv \sum_i\langle \hat{\sigma}^\gamma_i \rangle/N$ is the magnetization per site. Furthermore, for spin systems with spontaneously broken $\mathbb{Z}_2$-symmetry in the ground state, such as the TFIM, one has $m_z=0$ in the thermodynamic limit, and the mean distance measured along the $\hat{\sigma}^z$ direction is always $1/2$, regardless of the system parameters. 

The variance $\text{Var}(r_H)$ can be derived from the fundamental formula for the variance of a sum of random variables as 
\begin{align}
    \text{Var}(r_H) &= \text{Var}\left(\frac{1}{N}\sum_i X_i\right) \nonumber\\
    &= \frac{1}{N^2}\sum_i \text{Var}(X_i) + \frac{1}{N^2}\sum_{i \neq j} \text{Cov}(X_i, X_j),
\end{align}
where $X_i = (1 - s_i s'_i)/2$ is a binary variable (0 or 1), with $X_i^2 = X_i$ and $\langle X_i\rangle = (1-\langle \hat{\sigma}^\gamma_i \rangle^2)/2$. Its variance can be easily obtained as 
\begin{align}
    \text{Var}(X_i) = \langle X_i \rangle - \langle X_i \rangle^2 = \frac{1}{4}\left(1-\langle \hat{\sigma}^\gamma_i \rangle^4 \right).
    \label{eq:var_X}
\end{align}
In order to obtain the covariance $\text{Cov}(X_i, X_j) = \langle X_i X_j \rangle - \langle X_i \rangle \langle X_j \rangle$, we unfold the joint expectation as 
\begin{align}
    \langle X_i X_j \rangle &= \frac{1}{4} \left\langle (1 - s_i s'_i)(1 - s_j s'_j) \right\rangle \nonumber\\ 
    &= \frac{1}{4} \left( 1 - \langle \hat{\sigma}^\gamma_i \rangle^2 - \langle \hat{\sigma}^\gamma_j \rangle^2 + \langle \hat{\sigma}^\gamma_i \hat{\sigma}^\gamma_j \rangle^2 \right).
    \label{eq:XiXj}
\end{align}
The last equality has considered independent sampling of the two configurations, i.e., the four-spin term factorizes as $\langle s_i s_j s'_i s'_j \rangle = \langle s_i s_j \rangle \langle s'_i s'_j \rangle = \langle \hat{\sigma}^\gamma_i \hat{\sigma}^\gamma_j \rangle^2$. Subtracting Eqs.~\eqref{eq:var_X} and \eqref{eq:XiXj}, as well as the product of means $\langle X_i \rangle \langle X_j \rangle = \frac{1}{4}(1-\langle \hat{\sigma}^\gamma_i \rangle^2)(1-\langle \hat{\sigma}^\gamma_j \rangle^2)$, we obtain the total variance as follows:
\begin{align}
    \text{Var}(r_H) = &\frac{1}{4N^2} \sum_i \left(1-\langle \hat{\sigma}^\gamma_i \rangle^4 \right) \nonumber \\ 
    & + \frac{1}{4N^2} \sum_{i \neq j} \left( \langle\hat{\sigma}^\gamma_i\hat{\sigma}^\gamma_j\rangle^2 - \langle\hat{\sigma}^\gamma_i\rangle^2 \langle\hat{\sigma}^\gamma_j\rangle^2 \right).
    \label{eq:var_RH_general}
\end{align}
By introducing the connected correlation function $C_{r_{ij}} = \langle\hat{\sigma}^\gamma_i\hat{\sigma}^\gamma_j\rangle - \langle\hat{\sigma}^\gamma_i\rangle \langle\hat{\sigma}^\gamma_j\rangle$ and again assuming the homogeneity of the lattice, we have the final form of $\text{Var}(r_H)$ containing three terms:
\begin{align}
    \text{Var}(r_H) = \frac{1 - m_\gamma^4}{4N} + \frac{1}{4N^2} \sum_{i \neq j} \left( C_{r_{ij}}^2 + 2m_\gamma^2 C_{r_{ij}} \right).
    \label{eq:var_rH}
\end{align}
Thus, the geometry of statistical measurements in the configuration space is connected to conventional observables in the real space.

\subsection{Criticality of $\sqrt{\mathrm{Var}(r_H)}$}
\label{sec:criticality}

\begin{table*}[t!]
    \centering
    \begin{tabular}{l|c|c|c}
        \hline
        \hline
        \textbf{Model (Universality Class)} & \textbf{critical exponents} & \textbf{Applicability Check} & \textbf{Predicted Scaling} \\
        \hline
        2D Ising (2D Ising) & $\beta =1/8 , \nu =1 $ & $4\beta/\nu = 0.5 < 2$ (Yes) & $\sim L^{-1/4}$ \\
        \hline
        3D Ising (3D Ising) & $\beta =0.327,\nu =0.630$ & $4\beta/\nu \approx 2.07 < 3$ (Yes) & $\sim L^{-1.036}$ \\
        \hline
        2D kagome ice model (BKT) & $ \eta_1 =1/9 $& $2\eta(T_{c1}) = 2/9 < 2$ (Yes) & $\sim L^{-1/9}$ \\
        & $\eta_2= 1/4$ & $2\eta(T_{c2}) = 2/4 < 2$ (Yes) & $\sim L^{-1/4}$ \\
        \hline
        1D TFIM (2D Ising) &  & $4\beta/\nu = 0.5 < 1$ (Yes) & $\sim L^{-1/4}$ \\
        \hline
        2D TFIM (3D Ising) &  & $4\beta/\nu \approx 2.07 \not< 2$ (Marginal) & $\sim L^{-1.036}$ \\
        \hline
        \hline
    \end{tabular}
    \caption{Applicability check of the scaling law according to the criterion $4\beta/\nu < d$ ($2\eta < d$) for continuous (BKT) phase transition. The values of critical exponents for typical Ising spin models and phase transitions are from representative literature~\cite{Sachdev_2011,3D_ising, 3D_Ising_exponets, Wolf_kagome_1988, Suzuki1976}. }
    \label{tab:applicability}
\end{table*}

While the critical behavior of $\sqrt{\text{Var}(r_H)}$ has been numerically demonstrated in previous studies~\cite{Su2023, Su2024}, this subsection aims to provide its analytical scaling form at a conventional critical point. Considering the continuous phase transition of systems with Ising spins, such as the $Z_2$ symmetry breaking phase transition of TFIM, our derivation rests on the following key aspects:
\begin{enumerate}
    \item Zero magnetization: One has $\langle \hat{\sigma}^z_i \rangle = 0$ on each site; this leads to $m^2_z=0$. 
    \item Critical decay of the real-space correlation: For a $d$-dimensional lattice with a dynamic critical exponent $z$, the equal-time connected correlation function at criticality decays as $C_r \sim r^{-(d+z-2+\eta)}$. Here $r_{ij}=|i-j|$ is the distance between site $i$ and $j$; $\eta$ is the anomalous dimension.
    \item Hyperscaling relation: The critical exponents are related by the hyperscaling identity $2\beta = \nu(d+z-2+\eta)$. This relation is a fundamental consequence of renormalization group theory.
\end{enumerate}

In addition, the key technique of our derivation is to replace the discrete summation with integrals, that is, $\sum_{i \neq j}$ is approximated by an integral $N \int d^dr \sim L^d \int r^{d-1}dr$. Here, the discrete lattice with $L^d$ sites is approximated by the $d$-dimensional sphere, where $L$ is the lattice length. Then, we can obtain the finite-size scaling behavior of each term of Eq.~\eqref{eq:var_rH} as follows:
\begin{itemize}
    \item The first term is independent of system details as
    \begin{align}
        \label{eq:var_rH_1}
        \frac{1}{4N}(1- m_z^4) \sim L^{-d},
    \end{align}
    where we use the relation between the lattice size and length, i.e., $N=L^d$. 
    
    \item The correlation-squared term reads
    \begin{align}
        \label{eq:var_rH_2}
        &\frac{1}{4N^2} \sum_{i \neq j} C_{r_{ij}}^2 
        \sim
        \frac{1}{L^{2d}} \sum_{i \neq j} C_{r_{ij}}^2 \nonumber\\
        &\sim \frac{1}{L^{2d}} \left( L^d \int_{1}^L r^{-2(d+z-2+\eta)} r^{d-1} dr \right) \nonumber\\
        &\sim \frac{1}{L^{2d}} \left( L^d \int_{1}^L r^{d-4\beta/\nu-1} dr \right) \nonumber\\
        &\sim L^{-4\beta/\nu},
    \end{align}
    where the lattice constant is assumed as $1$. One may notice that the integral \( \int_1^L r^\kappa  dr \) exhibits distinct asymptotic behaviors depending on the exponent \( \kappa \): For \( \kappa > -1 \), the integral is dominated by its upper limit and scales as \( L^{\kappa+1} \); At \( \kappa = -1 \), the integral diverges logarithmically, \( \int_1^L r^{-1}  dr \sim \log L \); For \( \kappa < -1 \), the integral converges to a constant for large \( L \), as it is dominated by its lower limit. Therefore, the last line of Eq.~\eqref{eq:var_rH_2} requires $4\beta/\nu < d$. 
    \item The magnetization-correlation cross term disappears due to zero magnetization $m_z=0$ as
    \begin{align}
        \label{eq:var_rH_3}
        &\frac{1}{4N^2} \sum_{i \neq j} 2m_z^2C_{r_{ij}} = 0.
    \end{align}
\end{itemize}
To confirm the final scaling behavior of $\textrm{Var}(r_H)$, one needs to consider the competition between the first term [$\sim L^{-d}$ in Eq.~\eqref{eq:var_rH_1}] and the second term [$\sim L^{-4\beta/\nu}$ in Eq.~\eqref{eq:var_rH_2}]. This comparison leads to a direct criterion 
\begin{align}
    4\beta/\nu < d,
    \label{eq:criterion}
\end{align}
which is also in accord with the necessary condition for Eq.~\eqref{eq:var_rH_2}. As long as this applicable criterion is satisfied, the standard deviation of $P(r_H)$ at the critical point follows the universal scaling law
\begin{align}
    \label{eq:scaling_law}
    \sqrt{\textrm{Var}(r_H)} \sim L^{-2\beta/\nu},
\end{align}
and one can safely perform the finite-size scaling using the form
\begin{align}
    \sqrt{\mathrm{Var}(r_H)} L^{2\beta/\nu} = \mathcal{F}(tL^{1/\nu})
    \label{eq:scaling_con}
\end{align}
for conventional continuous phase transitions. Here $t$ is the parameter distance to the critical point. As shown in Table~\ref{tab:applicability}, this criterion is valid for typical phase transitions of classical Ising spin systems. The scaling law is also validated by the comprehensive numerical results for temperature-driven phase transitions in Ref.~\cite{Su2024}, where the numerically obtained critical exponent is close to the anomalous dimension $\eta$, which is equal to $2\beta/\nu$ in Eq.~\eqref{eq:scaling_law} when considering the hyperscaling relation for the relevant classical spin models.
This general derivation applies to both quantum ($z>0$) and classical ($z=0$) systems, provided that the corresponding exponents are employed.

\subsection{Quasi-long-range order}

The above derivation follows Landau's symmetry-breaking paradigm, explicitly considering the continuous phase transition between the long-range ordered and disordered phases. Nevertheless, the scaling behavior of $\sqrt{\textrm{Var}(r_H)}$ can be easily extended to the case of quasi-long-range order and BKT phase transition. The system with quasi-long-range order features:
\begin{enumerate}
    \item \textit{Zero magnetization:} The system has no local long-range order, i.e., $m_\gamma = 0$.
    \item \textit{Critical decay of real-space correlation:} The real-space correlations decay as a power law with distance: $C_r \sim r^{-\eta}$.
\end{enumerate}
Following similar derivation procedures in Sect.~\ref{sec:criticality}, one easily gets: i) the trivial term in Eq.~\eqref{eq:var_rH_1} remains unchanged; ii) the correlation-squared term in Eq.~\eqref{eq:var_rH_2} scales as $\sim L^{-2\eta}$; iii) the magnetization-correlation cross term in Eq.~\eqref{eq:var_rH_3} is zero. Therefore, we obtain the scaling law 
\begin{align}
    \sqrt{\textrm{Var}(r_H)}\sim L^{-\eta}
\end{align}
for the phase transition between the quasi-long-range ordered and the disordered phase. This explains the numerically obtained critical exponents for BKT phase transitions of the kagome ice model in Ref.~\cite{Su2024}, where a similar criterion $2\eta<d$ is satisfied.

\subsection{Connection to the static structure factor}
\label{sec:connection_to_S_q}

The analytical connection in Eq.~\eqref{eq:var_rH} between the variance of configuration-space distances and the sum of squared two-point real-space correlations naturally implies an equivalent perspective in momentum space. For a system with $N$ sites, the discrete Parseval's theorem establishes a direct identity between the sum of squared correlations in real space and the sum of the squared intensities of the structure factor in momentum space as 
\begin{equation}
    \sum_{i,j} |C(\mathbf{r}_i - \mathbf{r}_j)|^2 =  \sum_{\mathbf{q}} |S(\mathbf{q})|^2,
    \label{eq:parseval}
\end{equation}
where the static structure factor $S(\mathbf{q}) = \sum_r e^{-i\mathbf{q}\cdot\mathbf{r}} C(\mathbf{r})$  is the Fourier transform of the correlation function. Combining this identity with previous discussions, we arrive at a fundamental relationship
(valid under the $m_\gamma=0$ assumption)
\begin{equation}
    \mathrm{Var}(r_H) \propto \frac{1}{N^2} \sum_{\mathbf{q}} |S(\mathbf{q})|^2,
    \label{eq:var_sq_relation}
\end{equation}
as long as the applicable criterion in Eq.~\eqref{eq:criterion} holds. It follows that the variance of the Hamming distance is proportional to the integrated intensity of the static structure factor, normalized by $N^2$, throughout the Brillouin zone.

This relationship provides a clear physical mechanism for how our observable detects magnetic ordering. A phase transition into a magnetically ordered state—whether ferromagnetic, antiferromagnetic, or incommensurate—is characterized by the emergence of sharp Bragg peaks in $S(\mathbf{q})$ at the specific ordering wavevectors $\mathbf{Q}$. These peaks produce a large, system-size-dependent intensity in the integrated structure factor $\sum_{\mathbf{q}} |S(\mathbf{q})|^2$, which is directly proportional to $N^{2}\mathrm{Var}(r_H)$. Consequently, the onset of long-range order is signaled by a distinct finite-size scaling of $\mathrm{Var}(r_H)$. While a complete analysis of $S(\mathbf{q})$ provides detailed information about the nature of the order, $\mathrm{Var}(r_H)$ offers a computationally simple, scalar alternative that is sensitive to the emergence of periodic order at any wavevector, without prior knowledge of$\mathbf{Q}$.

\section{Numerical validation of configuration-space criticality}
\label{sec:numerics}

Having established the analytical framework, which also successfully explains numerical results of finite-temperature phase transitions in a series of classical spin systems~\cite{Su2024}, we now present numerical results on TFIM, further validating the analytical framework for the quantum phase transition. 
We employ the stochastic series expansion (SSE) quantum Monte Carlo method to approximately sample the configurations that follow the $P(s)=\langle s|\Psi_0 \rangle$ with the ground state $|\Psi_0\rangle$, and further obtain the estimated $P(r_H)$ of pairwise distances between sampled configurations. In this section, we simulate the TFIM Hamiltonian in Eq.~\eqref{eq:H_TFIM} using an efficient quantum cluster algorithm~\cite{Sandvik_SSE_TFIM} along the $\hat{\sigma}^z$ basis, and the ground state is approximated by the simulation at a low temperature $T=1/L$~\cite{Sandvik2010}, with periodic boundary conditions (PBC). Other numerical details, especially how configurations and distances are collected, are introduced in Appendix~\ref {app:num_details}.

\begin{figure}[!t]
    \centering
    \includegraphics[width=\linewidth]{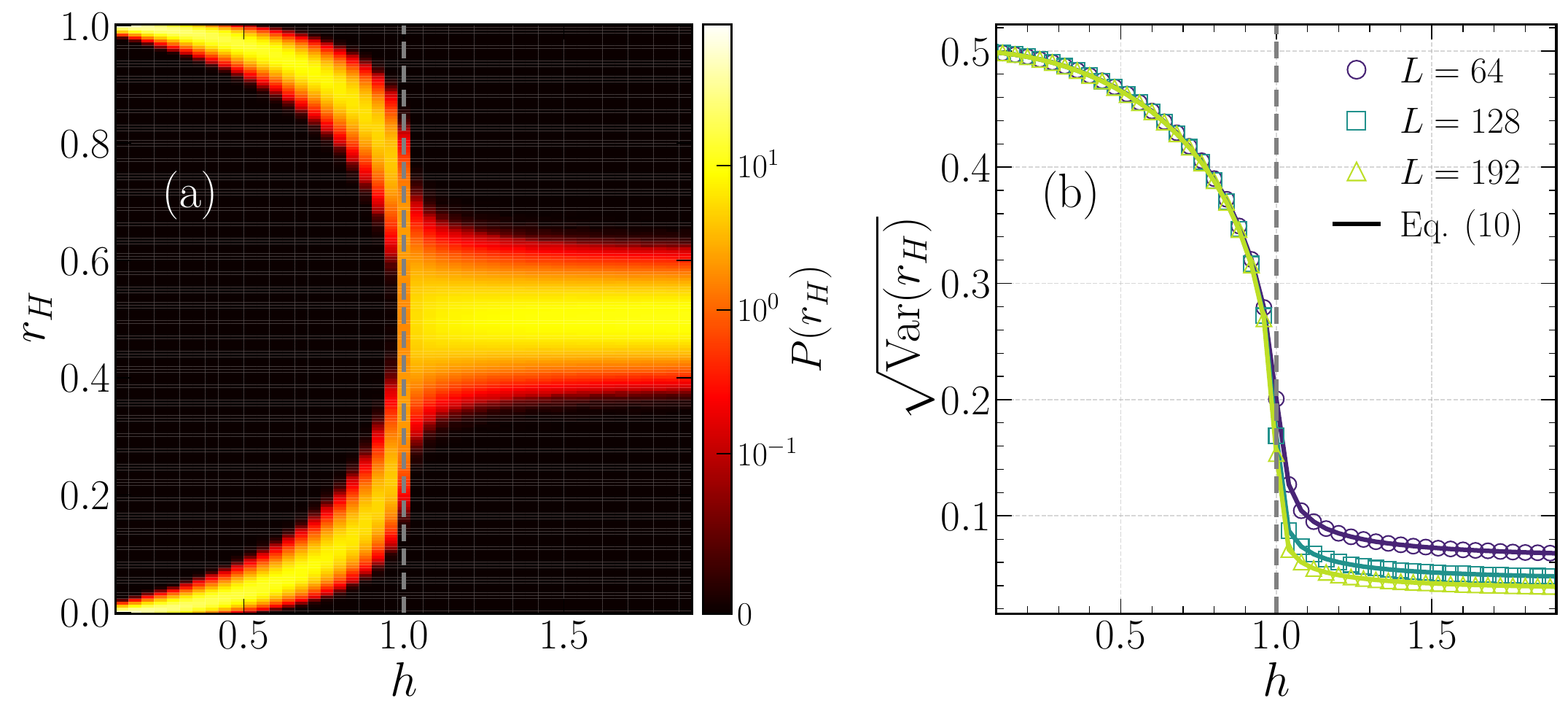} 
    \caption{(a) The evolution of probability distribution $P(r_{H})$ as a function of the transverse field $h$, for 1D TFIM with $L=192$. (b) The standard deviation $\sqrt{\textrm{Var}(r_H)}$ extracted from data statistics, depicted by empty markers; and the theoretically estimated value using real-space observables according to Eq.~\eqref{eq:var_rH}, depicted by solid lines. In both panels, the vertical dashed line depicts the critical point $h_c=1$. }
    \label{fig:Pr_1D}
\end{figure}

The TFIM Hamiltonian reads
\begin{align}
    H_\mathrm{TFIM} = -J\sum_{\langle ij\rangle} \hat{\sigma}^z_i \hat{\sigma}^z_j - h\sum_i \hat{\sigma}^x_i,
    \label{eq:H_TFIM}
\end{align}
where the sum $\langle ij\rangle$ runs over the nearest-neighbor sites and ferromagnetic interaction $J = 1$ is set as the energy scale. At zero temperature, $T=0$, the system exhibits a quantum phase transition at the critical transverse field $h=h_c$ in one and two dimensions. For the zero transverse field, $h=0$, the system degenerates to the classical Ising model and undergoes a finite-temperature phase transition at $T=T_c$ in two and three dimensions. The finite-temperature phase transition in $d+1$-dimensional classical Ising systems shares the same universality class as the quantum phase transition in $d$-dimensional TFIM~\cite{Suzuki1976}.

\subsection{1D TFIM}

The feasibility of our proposal relies on the fundamental assumption that the statistical geometry in the Hilbert space contains the essential information of phases and phase transitions. In this work, the statistical geometry is constructed by pairwise distances between configurations, and the very basic statistical measurement is the plain probability distribution of distances. 

In Fig.~\ref{fig:Pr_1D}(a), we visualize how the probability distribution $P(r_{H})$ evolves with the transverse field strength $h$ for the quantum phase transition in 1D TFIM. At $h=0$, the system resides in one of two degenerate ferromagnetic ground states. Consequently, the distance between any two sampled configurations can only be 0 (if they originate from the same ground state) or 1 (if from different ones), appearing as two bright branches in the plot. As the transverse field $h$ increases, two branches spread gradually due to quantum fluctuations. As the system passes through the quantum critical point into the paramagnetic phase, long-range correlations vanish, and the configurations become random. The two branches eventually merge into a single, narrowing Gaussian-like distribution centered at $\langle r_H\rangle=0.5$.

\begin{figure}[!t]
    \centering
    \includegraphics[width=\linewidth]{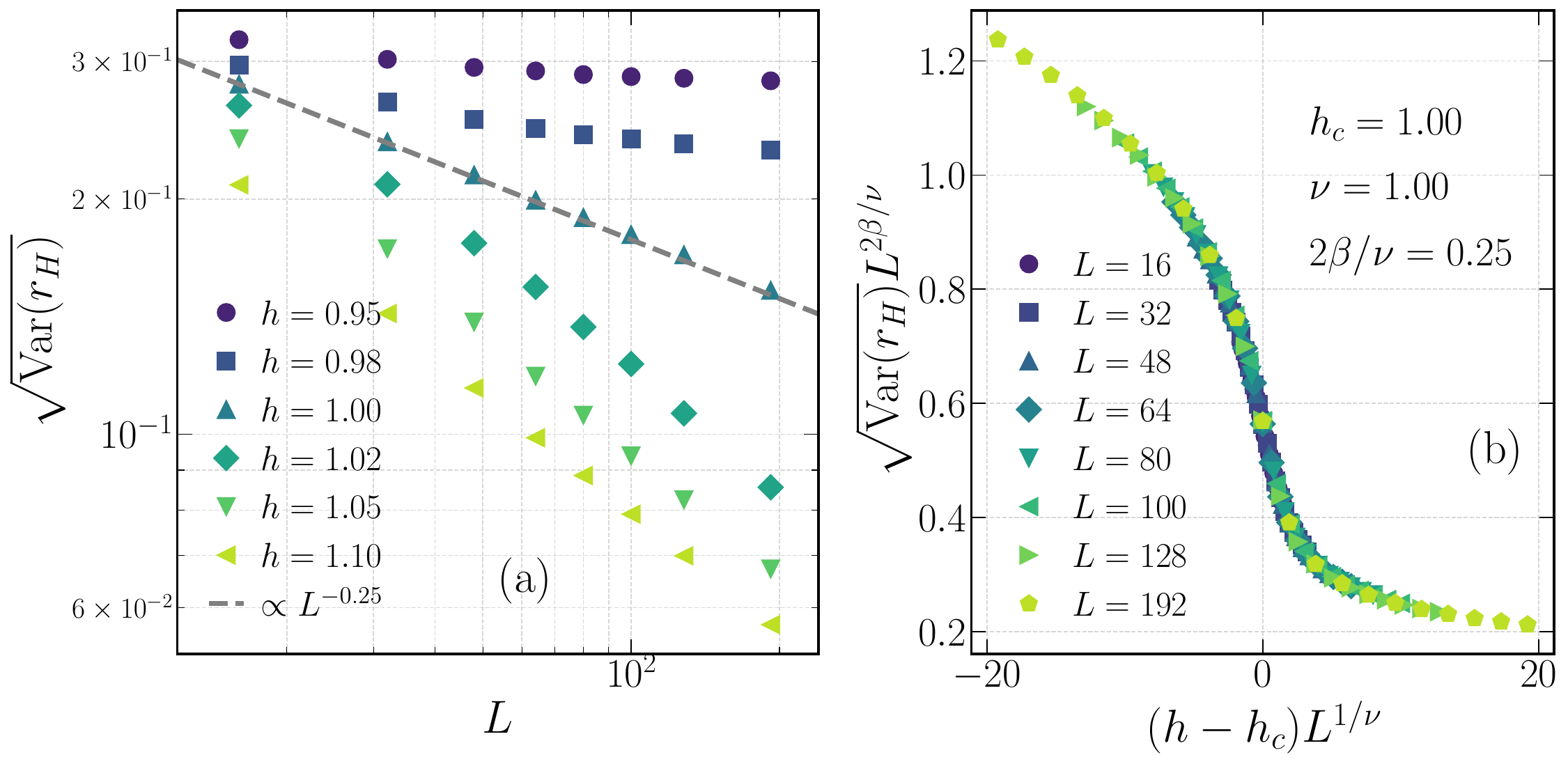} 
    \caption{Finite-size scaling analysis for $\sqrt{\mathrm{Var}(r_H)}$ of 1D TFIM. (a) The double-logarithmic plot of $\sqrt{\mathrm{Var}(r_H)}$ versus system size $L$ for several values of the transverse field $h$ near the critical point. The solid line depicts the best power-law fitting $\sim L^{-1/4}$ at the critical point. (b) Data collapse according to the scaling ansatz in Eq.~\eqref{eq:scaling_con}. 
    The optimal parameters obtained from minimizing the collapse error [Here and after, the data collapse follows the procedure in Appendix~\ref{app:scaling_details}], and they agree with the known values for the 2D Ising universality class ($h_c=1$, $\nu=1$, $2\beta/\nu=1/4$).  }
    \label{fig:1d_fss}
\end{figure}

While one can intuitively read different phases from the probability distribution $P(r_{H})$, its uncertainty further proposes the phase transition by a sharp jump around the critical $h_c=1.0$, as displayed in Fig.~\ref{fig:Pr_1D}(b). We also compare the extracted $\sqrt{\mathrm{Var}(r_H)}$ directly from statistics in configuration space with that reconstructed using real-space observables according to Eq.~\eqref{eq:var_rH}, where the excellent agreement verifies our analytical derivation. For all lines with different system sizes, the rapid drop around $h_c=1$ and the orderly finite-size dependence indicate their critical scaling behavior.  

For 1D TFIM, the criterion $4\beta/\nu < d$ is satisfied, as shown in Table~\ref{tab:applicability}; we expect $\sqrt{\mathrm{Var}(r_H)} \sim L^{-2\beta/\nu} = L^{-1/4}$ at the critical point, according to our theoretical prediction in Eq.~\eqref{eq:scaling_law}. This is perfectly verified by the SSE simulation results, as shown in Fig.~\ref{fig:1d_fss}(a). Furthermore, we perform a data collapse procedure according to the scaling ansatz in Eq.~\eqref{eq:scaling_con}; all data points for different $L$ and $h$  collapse onto a single smooth curve, as shown in Fig.~\ref{fig:1d_fss}(b). This perfect match between the theoretical prediction and the numerical results provides strong validation of our new scaling law on the configuration-space statistical geometry.

\subsection{2D TFIM}

\begin{figure}[!b]
    \centering
    \includegraphics[width=\linewidth]{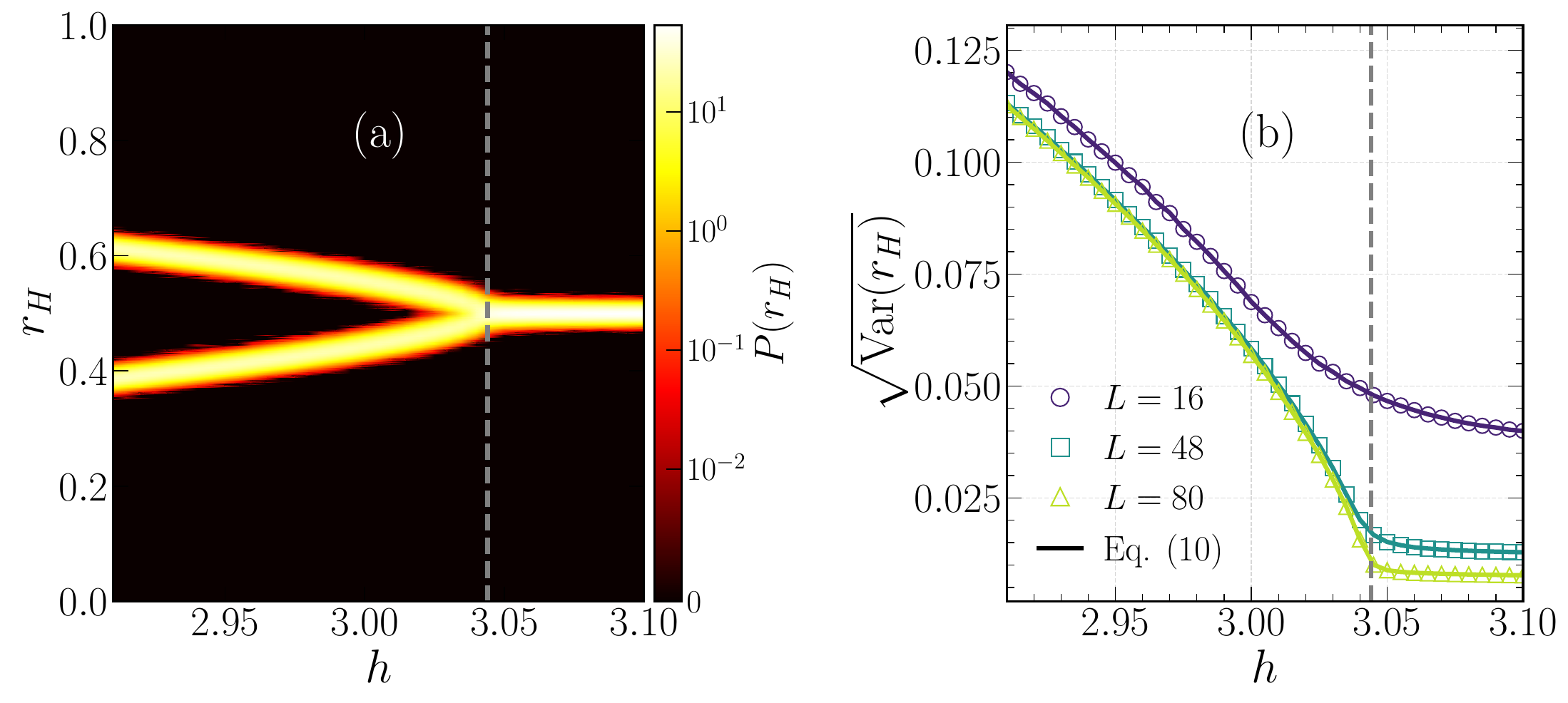} 
    \caption{(a) The evolution of probability distribution $P(r_{H})$ as a function of the transverse field $h$, for 2D TFIM with $L=80$. (b) The standard deviation $\sqrt{\textrm{Var}(r_H)}$ extracted from data statistics, depicted by empty markers; and the theoretically estimated value using real-space observables according to Eq.~\eqref{eq:var_rH}, depicted by solid lines. In both panels, the vertical dashed line depicts the critical point $h_c=3.044$. }
    \label{fig:Pr_2D}
\end{figure}

The 2D TFIM exhibits a quantum phase transition that belongs to the 3D classical Ising universality class, with $4\beta/\nu \approx 2.072$, which is slightly beyond the lattice dimension $d=2$. This case presents a subtle theoretical challenge to the scaling law in configuration space, as the applicable criterion in Eq.~\eqref{eq:criterion} is marginally violated. Nevertheless, it is necessary to conduct the numerical experiment for the 2D TFIM to check whether and to what extent the geometry of the configuration space captures the criticality.

\begin{figure}[!t]
    \centering
    \includegraphics[width=\linewidth]{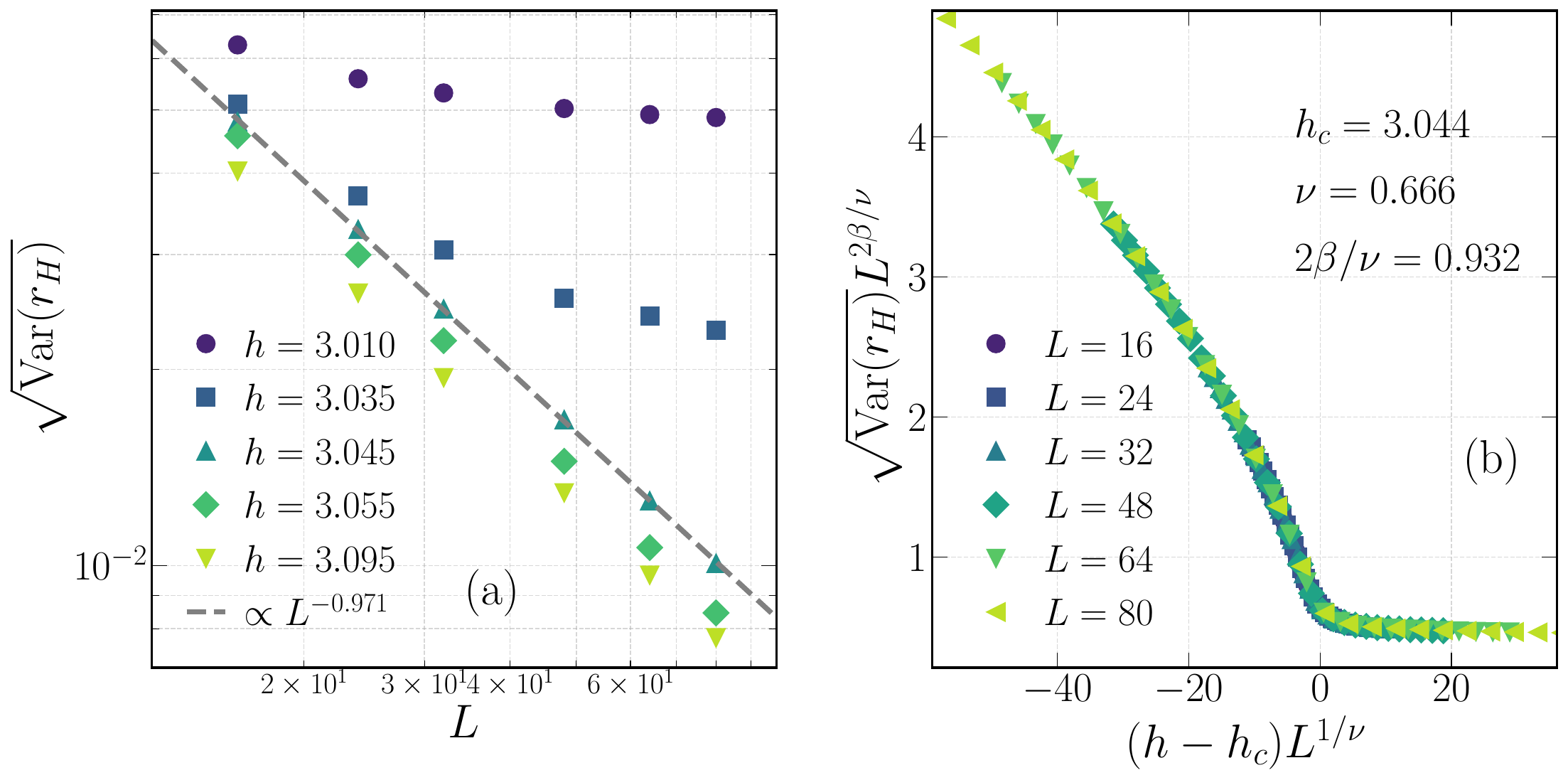} 
    \caption{Finite-size scaling analysis for $\sqrt{\mathrm{Var}(r_H)}$ of 2D TFIM. (a) The double-logarithmic plot of $\sqrt{\mathrm{Var}(r_H)}$ versus system size $L$ for several values of the transverse field $h$ near the critical point. The solid line depicts the best power-law fitting $\sim L^{-0.971}$ at $h=3.045$, which is the closest data point to the known $h_c$. (b) Data collapse according to the scaling ansatz in Eq.~\eqref{eq:scaling_con}. The optimal parameters obtained from minimizing the collapse error; the numerical value obtained critical point agrees well with the known values for the 3D Ising universality class ($h_c=3.044$);  $\nu=0.666$ and $2\beta/\nu=0.932$ are slightly away from the theoretical value ($\nu=0.630$,  $2\beta/\nu=1.036$).}
    \label{fig:2d_fss}
\end{figure}

Similarly to the 1D case, the numerically obtained distribution probability $P(r_H)$ strongly suggests two phases and a phase transition, as shown in Fig.~\ref{fig:Pr_2D}(a). The match between the statistical measurement in the configuration space and the conventional observables in the real space for the 2D case is testified in Fig.~\ref{fig:Pr_2D}(b), where lines of different system sizes again indicate possible universal scaling behavior.  

\begin{figure}[!b]
    \centering
    \includegraphics[width=\linewidth]{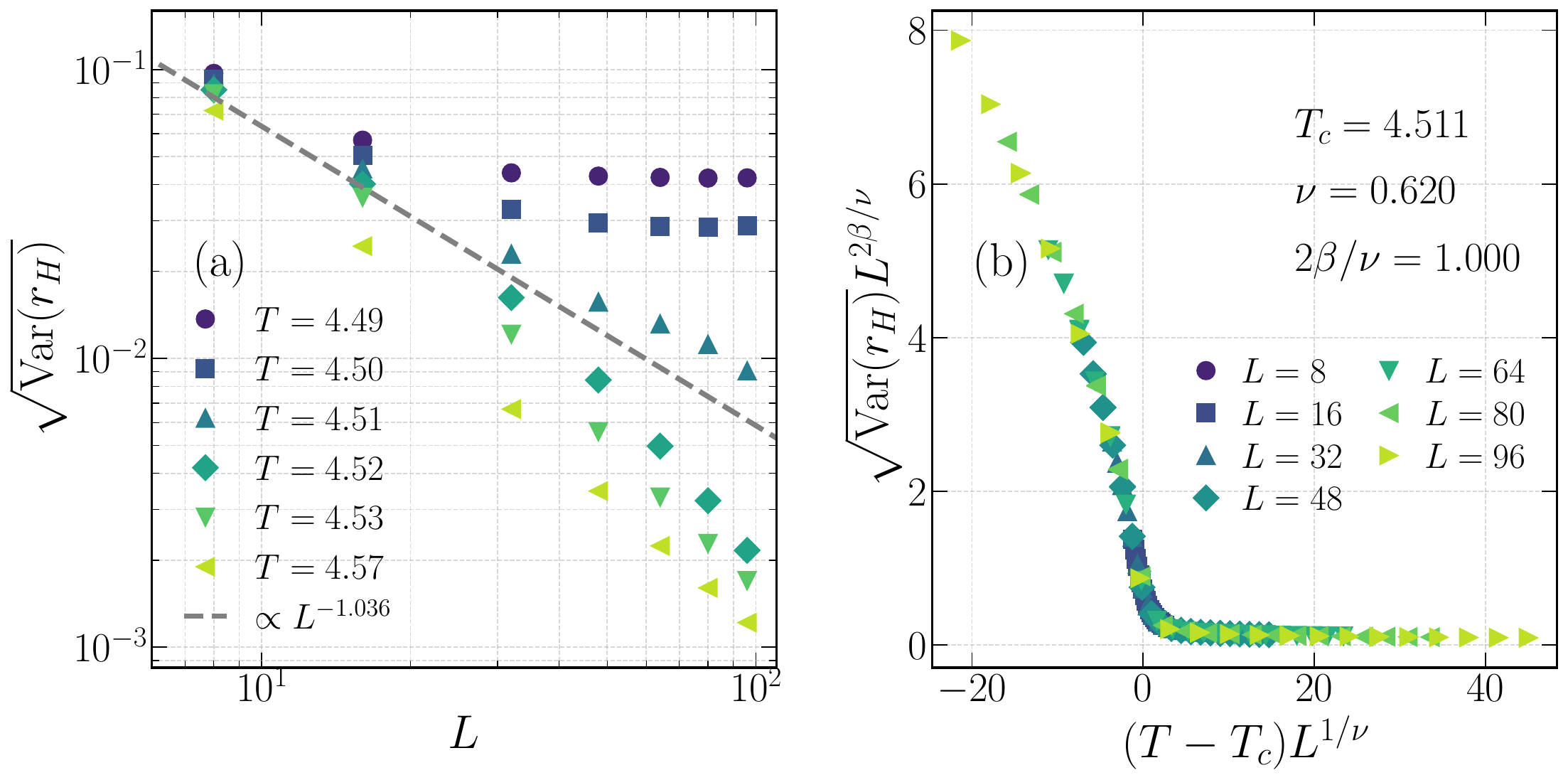} 
    \caption{Finite-size scaling analysis for $\sqrt{\mathrm{Var}(r_H)}$ of 3D classical Ising model, i.e., TFIM with $h=0$. (a) The double-logarithmic plot of $\sqrt{\mathrm{Var}(r_H)}$ versus system size $L$ for several values of temperatures $T$ near the critical point. The solid line depicts the predicted critical decay $\sim L^{-1.036}$. (b) Data collapse according to the scaling ansatz in Eq.~\eqref{eq:scaling_con}. The optimal parameters are obtained by minimizing the collapse error. }
    \label{fig:3d_Ising_fss}
\end{figure}

A more rigorous finite-size analysis of $\sqrt{\mathrm{Var}(r_H)}$ is displayed in Fig.~\ref{fig:2d_fss}(a), where one can find an evident algebraic decay near the critical point. However, the decay ratio is slightly different from the predicted value $2\beta/\nu$. When performing the best data collapse following Eq.~\eqref{eq:scaling_con}, we still catch the critical point well, even though the numerically obtained critical exponents are not as perfect as for the 1D TFIM. Actually, since the applicable criterion in Eq.~\eqref{eq:criterion} is marginally violated, the numerically obtained critical exponent can be either $2\beta/\nu=1.036$ or $d/2=1$, which are indistinguishable via finite-size numerical results. 

We also make a direct comparison, considering the temperature-driven phase transition of the 3D classical Ising model, which belongs to the same universality class as the 2D TFIM. While all critical exponents are the same, the key difference between the two cases lies in the lattice dimension $d$. Therefore, the 3D Ising model satisfies the criterion in Eq.~\eqref{eq:criterion}, with $4\beta/\nu \approx 2.072 < 3$. As shown in Fig.~\ref{fig:3d_Ising_fss}, in addition to the accurate critical point, the critical exponents extracted from $\sqrt{\mathrm{Var}(r_H)}$ are much closer to theoretical predictions, further validating our theoretical framework.

\subsection{Control Experiment: Orthogonal Basis Analysis}
\label{sec:x_basis}

\begin{figure}[!b]
    \centering
    \includegraphics[width=\linewidth]{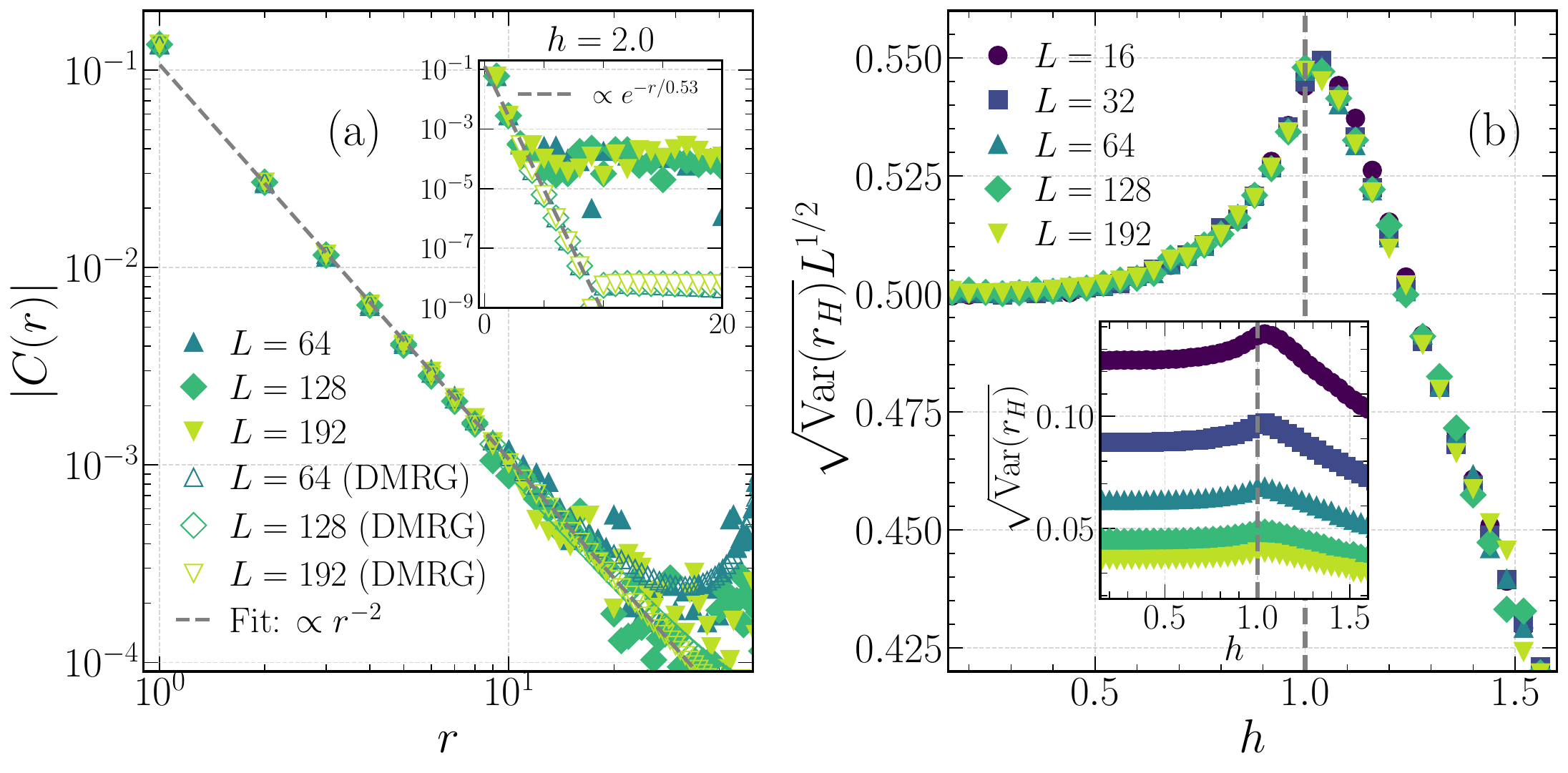} 
    \caption{Control experiment for measurements along $\hat{\sigma}^x$ basis of 1D TFIM. (a) The power-law decay of the absolute value of the real-space correlation $|C_r|$ at the critical point $h=1$, where filled/empty samples represent QMC/DMRG results, respectively. The inset of (a) shows the exponential decay of $|C_r|$ away from the critical point. Simulations from both methods are performed under PBC. A slight mismatch in the inset of panel (a) is due to i) the statistical error background of QMC and ii) the SSE algorithm is not (able to be) performed at exact zero temperature. (b) The data collapse of $\sqrt{\mathrm{Var}(r_H)}$ following the system-size dependence in Eq.~\eqref{eq:Var_rH_x}. The vertical dashed line depicts the critical point $h_c = 1$. The inset of (b) displays the raw $\sqrt{\mathrm{Var}(r_H)}$ versus $h$ for different $L$; it shares the same legend as the main panel.}
    \label{fig:x_basis_analysis}
\end{figure}

This subsection conducts a crucial control experiment by analyzing the distance along the $\hat{\sigma}^x$ basis. In this case, the magnetization $m_x$ is nonzero, and the real-space correlation function along $\sigma_x$ direction does not capture a spontaneous symmetry breaking [See Appendix~\ref{app:x_basis_analysis}]. We expect that $\sqrt{\mathrm{Var}(r_H)}$ loses its criticality as local probes in the $\hat{\sigma}^x$ basis miss the full critical information. 

For numerical convenience, we perform the following two-step operations on Pauli matrices: i) rotate the spin by a global angle $\pi$, $\hat{\sigma}^z \to -\hat{\sigma}^x,\quad\hat{\sigma}^x \to \hat{\sigma}^z$; and ii) replace $\hat{\sigma}^x$ with raising/lowering operators, $\hat{\sigma}^x = \hat{\sigma}^+ + \hat{\sigma}^-$. Then we get a hard-core boson Hamiltonian with single-boson and pair hopping:
\begin{align}
    H_\mathrm{HB} = -J \sum_{ij} \left( \hat{\sigma}_i^+ \hat{\sigma}_{j}^+ + \hat{\sigma}_i^+ \hat{\sigma}_{j}^- + h.c. \right) - h \sum_i \sigma_i^z
    \label{eq:H_rotated}
\end{align}
We solve this Hamiltonian by performing SSE QMC simulations with a directed loop algorithm~\cite{Sandvik_SSE_DL}, and again measure the distance along the $\hat{\sigma}^z$ direction, which equals measurements along $\hat{\sigma}^x$ for the initial TFIM Hamiltonian in Eq.~\eqref{eq:H_TFIM}. 

Although the hard-core boson Hamiltonian $H_\mathrm{HB}$ supports exactly the same phase transition as the 1D TFIM, $\sqrt{\mathrm{Var}(r_H)}$ measured in different basis behaves differently. 
Let us start our analysis by checking the three terms in Eq.~\eqref{eq:var_rH}, which describe the explicit connection between $\mathrm{Var}(r_H)$ and real-space observables. 
\begin{itemize}
    \item The first term remains unchanged, following the same power-law decay $\sim L^{-d}$ in Eq.~\eqref{eq:var_rH_1}, as long as $m_x<1$. This is always true in the finite $h$ regime we are interested in.
    \item The second term depends on the real-space correlation function along the $\hat{\sigma}^x$ direction, which decays algebraically ($\sim r^{-2}$) at the critical point and exponentially elsewhere~\cite{Pfeuty1970}. As displayed in Fig.~\ref{fig:x_basis_analysis}(a), the correlation decay is confirmed by our numerical results, obtained from both QMC and density matrix renormalization group (DMRG) method~\cite{White1992,White1993}. 
    In this case, the integral $\int_{1}^L C_r^2 r^{d-1} dr$ is dominated by its lower limit and converges to a constant, i.e., the second term scales as $L^{-d}$ for arbitrary $h$.
    \item The integral $\int_{1}^L m_x C_r r^{d-1} dr$ of the third term converges to a constant as long as $m_x$ is finite. Therefore, the third term features the scaling law $L^{-d}$ for any finite $h$.  
\end{itemize}     
Therefore, a trivial form dominates the final scaling law of the variance, and the standard deviation has the system-length dependence as
\begin{align}
    \sqrt{\mathrm{Var}(r_H)} \sim L^{-d/2}.
    \label{eq:Var_rH_x}
\end{align}
Unlike 2D TFIM, where $\sqrt{\mathrm{Var}(r_H)}$ clearly distinguishes the ordered and disordered phase, one should note that Eq.~\eqref{eq:Var_rH_x} is not a scaling law for criticality; instead, it provides a non-critical rescaling form for the entire parameter region. In other words, $\sqrt{\mathrm{Var}(r_H)}$ fails to catch the full criticality when measured in $\hat{\sigma}^x$ basis, although it features a broad peak at the critical point, as shown in Fig.~\ref{fig:x_basis_analysis}(b).

\section{Non-local probes in configuration-space}
\label{sec:nonlocal_probes}

Up to now, we have demonstrated how the phase transition and its universal criticality can be extracted from the probability distribution of configuration-space distances. Specifically, its standard deviation $\sqrt{\mathrm{Var}(r_H)}$ behaves as an order parameter under the particular applicable criterion. Our analytic derivation not only unveils the underlying physics of previous numerical successes along the line~\cite{Su2023,Su2024,Yi2022}, but also provides further physical insight into why ML approaches via processing configuration space distances can capture the critical physics~\cite{Facco2017,Rodriguez-Nieva2019,santos1,Santos2021prxq}.

However, the analytical link between $\mathrm{Var}(r_H)$ and the real-space correlations also restricts its application; that is, it depends on the measurement basis (as discussed in Sec.~\ref{sec:x_basis}) and provides no more information than the local probes in real space. A crucial question naturally arises: how can one extract critical information beyond real-space local probes from the configuration-space distance distribution $P(r_H)$? In this section, we provide two simple examples demonstrating how $P(r_H)$ indeed contains more information than real-space local probes.

\subsection{Fisher information as basis independent measurement}
\label{sec:NL_fisher_information}

In this subsection, we analyze the correlation between distributions of different parameters in the sense of information geometry~\cite{Amari2000,Nielsen2020}. In practice, one can establish a statistical manifold $\mathcal{M}$ by parametrizing the distribution of lattice configurations $P(r_H;h)$ with the transverse field $h$, where each point on $\mathcal{M}$ uniquely corresponds to a quantum state of the system. Intuitively, criticality is geometrically encoded in this manifold: at the quantum phase transition, scale-invariant fluctuations induce a singularity in the Riemannian curvature, marked by the divergence of the Fisher information metric~\cite{Amari2000,Nielsen2020,Zanardi2008}. For the 1D manifold with a single continuous parameter $h$, the Fisher information metric degenerates to the scalar Fisher information $\mathcal{I}(h)$, which is defined as 
\begin{align}
    \mathcal{I}(h) = \sum_{r_H} P(r_H;h) \left( \frac{\partial \log P(r_H;h)}{\partial h} \right)^2
    \label{eq:FI}
\end{align}
for the discrete distance $r_H$.

\begin{figure}[!t]
    \centering
    \includegraphics[width=\linewidth]{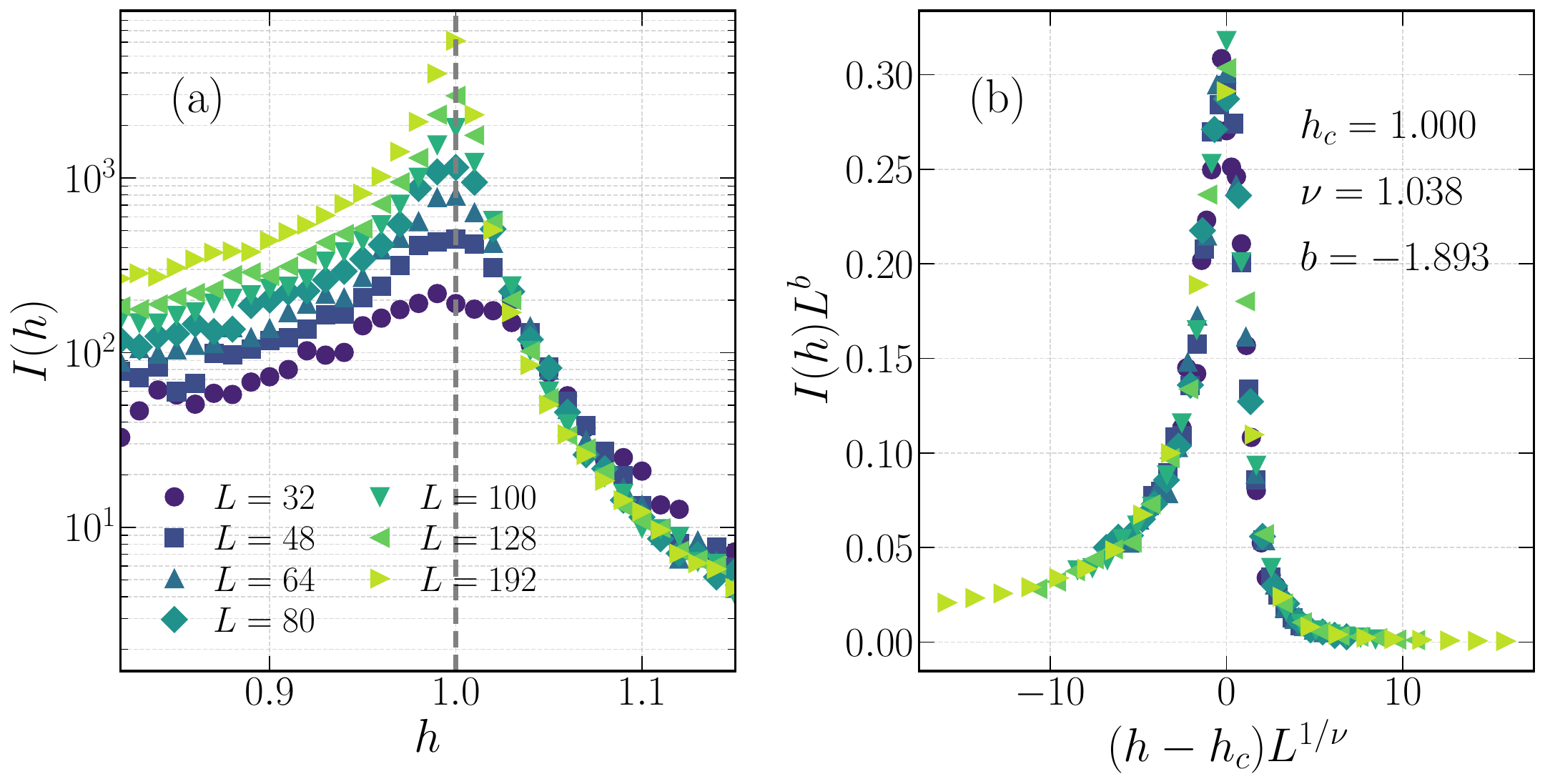} 
    \caption{Criticality of 1D TFIM captured by the information geometry measurement in $\sigma^z$ basis. (a) Fisher information $\mathcal{I}(h)$ versus $h$ for different system length $L$. The vertical dashed line depicts the critical point $h_c=1$. (b) The data collapse according to the scaling ansatz in Eq.~\eqref{eq:scaling_I}. The numerically obtained $h_c$ and $\nu$ for the best data collapse agree well with the known values.}
    \label{fig:1d_fss_FI}
\end{figure}

As shown in Fig.\ref{fig:1d_fss_FI}(a), the Fisher information measured along $\sigma^z$ basis of 1D TFIM features a singularity at the critical point: $\mathcal{I}(h)$ forms evident peaks for all system sizes; as $L$ increases, the peak position approaches $h_c=1$, and the peak height grows dramatically. This singularity encourages us to perform the data collapse following the general scaling ansatz:
\begin{align}
    \mathcal{I}(h) L^{b} = \mathcal{G}\left(|h-h_c|L^{1/\nu}\right),
    \label{eq:scaling_I}
\end{align}
where $b$ is the specific critical exponent of $\mathcal{I}(h)$. The quality of this brute-force data collapse is quite good, with all data points of different $h$ and $L$ collapsing into a narrow curve that exhibits a sharp peak at the critical point, as shown in Fig.~\ref{fig:1d_fss_FI}(b). The optimal critical parameters $h_c$ and $\nu$ for the best data collapse also match the knowledge of 1D TFIM.  

\begin{figure}[!t]
    \centering
    \includegraphics[width=\linewidth]{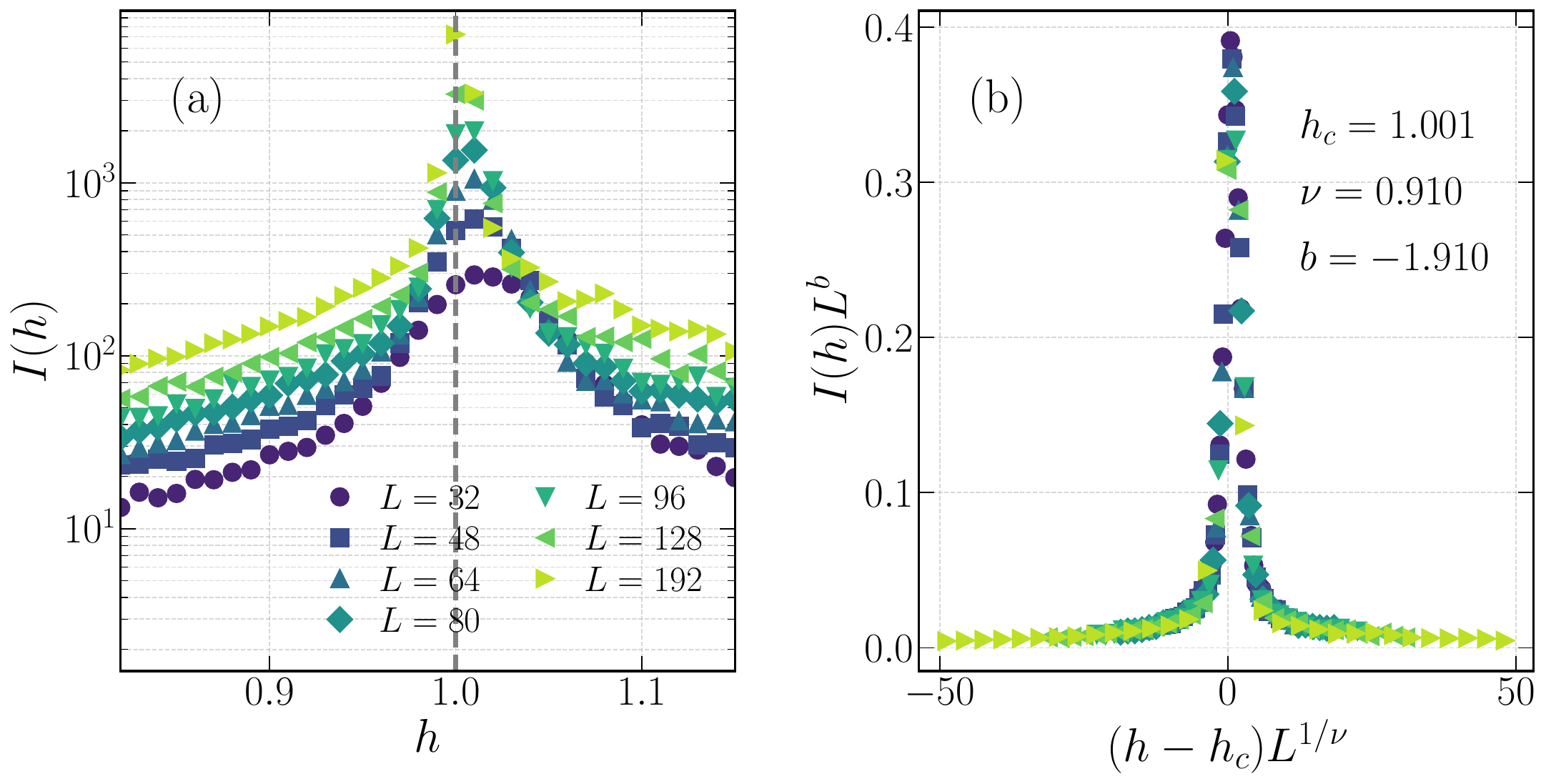} 
    \caption{Criticality of the 1D hard-core boson Hamiltonian in Eq.~\eqref{eq:H_rotated} captured by the information geometry measurement. (a) Fisher information $\mathcal{I}(h)$ versus $h$ for different system length $L$. The vertical dashed line depicts the critical point $h_c=1$. (b) The data collapse according to the scaling ansatz in Eq.~\eqref{eq:scaling_I}. The numerically obtained $h_c$ and $\nu$ for the best data collapse agree well with the known results.}
    \label{fig:1d_fss_HB}
\end{figure}

Unlike the standard deviation $\sqrt{\mathrm{Var}(r_H)}$ explicitly related to local real-space observables, $\mathcal{I}(h)$ measures the singularity in the manifold of distribution probabilities of distances in configuration space. In the negative aspect, we have not yet built the connection between $\mathcal{I}(h)$ and well-known critical physics, and this fact leaves an unknown critical exponent $b$ in Eq.~\eqref{eq:scaling_I}. On the positive side, $\mathcal{I}(h)$ may play the role of a brand new global measurement in the configuration space that is independent of the choice of the basis. Therefore, following the setting in Sec.~\ref{sec:x_basis}, we examine the Fisher information along the basis orthogonal to the order parameter of TFIM. The numerical results in Fig.~\ref{fig:1d_fss_HB} verify our intuitive guess: $\mathcal{I}(h)$ of the hard-core boson Hamiltonian~\eqref{eq:H_rotated} features quantitatively the same behavior as in Fig.~\ref{fig:1d_fss_FI} and captures the criticality of the phase transition. Considering the moderate data quality of $\mathcal{I}(h)$, the optimal $h_c$ and $\nu$ agree with the well-known results, and the critical exponent $b=-1.910$ is also numerically close to that obtained for TFIM in Fig.~\ref{fig:1d_fss_FI}. Therefore, we strongly suspect that the Fisher information along different measuring bases shares the same universal scaling behavior. 

\subsection{The symmetry-protected topological order and parity index }
\label{sec:NL_parity_index}

The symmetry-protected topological (SPT) phase is a canonical example of a state that cannot be described by local order parameters and lies beyond the LGW paradigm. It is interesting to check whether the measurements based on $P(r_H)$ can tackle SPT phases and phase transitions. To address this question, we turn to the spin-1/2 Heisenberg chain with alternating antiferromagnetic bonds, whose noninteracting version degenerates to the Su-Schrieffer–Heeger (SSH) model~\cite{ssh1968,Verresen2017} supporting the simplest SPT phase. The Hamiltonian is given by:
\begin{equation}
    H_{SSH} = J_1 \sum_{i \in \text{odd}} \mathbf{S}_i \cdot \mathbf{S}_{i+1} + J_2 \sum_{i \in \text{even}} \mathbf{S}_i \cdot \mathbf{S}_{i+1},
    \label{eq:ssh_heisenberg}
\end{equation}
where $\mathbf{S}_i=(S^x_i, S^y_i, S^z_i)$ is the Heisenberg spin on site $i$. The bulk states on either side of the Heisenberg limit $J_2/J_1=1$ are dimerized insulators, indicating the absence of a conventional symmetry-breaking transition. In terms of symmetry-protected topological (SPT) order, however, this quantum many-body model with short-range interactions is expected to exhibit the same topological physics as the non-interacting SSH model. Specifically, the system is topologically trivial for $J_2/J_1 < 1$ and non-trivial for $>1$, with a topological phase transition at $J_2/J_1 = 1$. The topologically non-trivial phase is characterized by the bulk-edge correspondence: a non-zero topological invariant defined as the integral of wavefunctions across the Brillouin zone under PBC, and the zero-energy edge excitations under open boundary conditions (OBC).

\begin{figure}[b!]
    \centering
    \includegraphics[width=\linewidth]{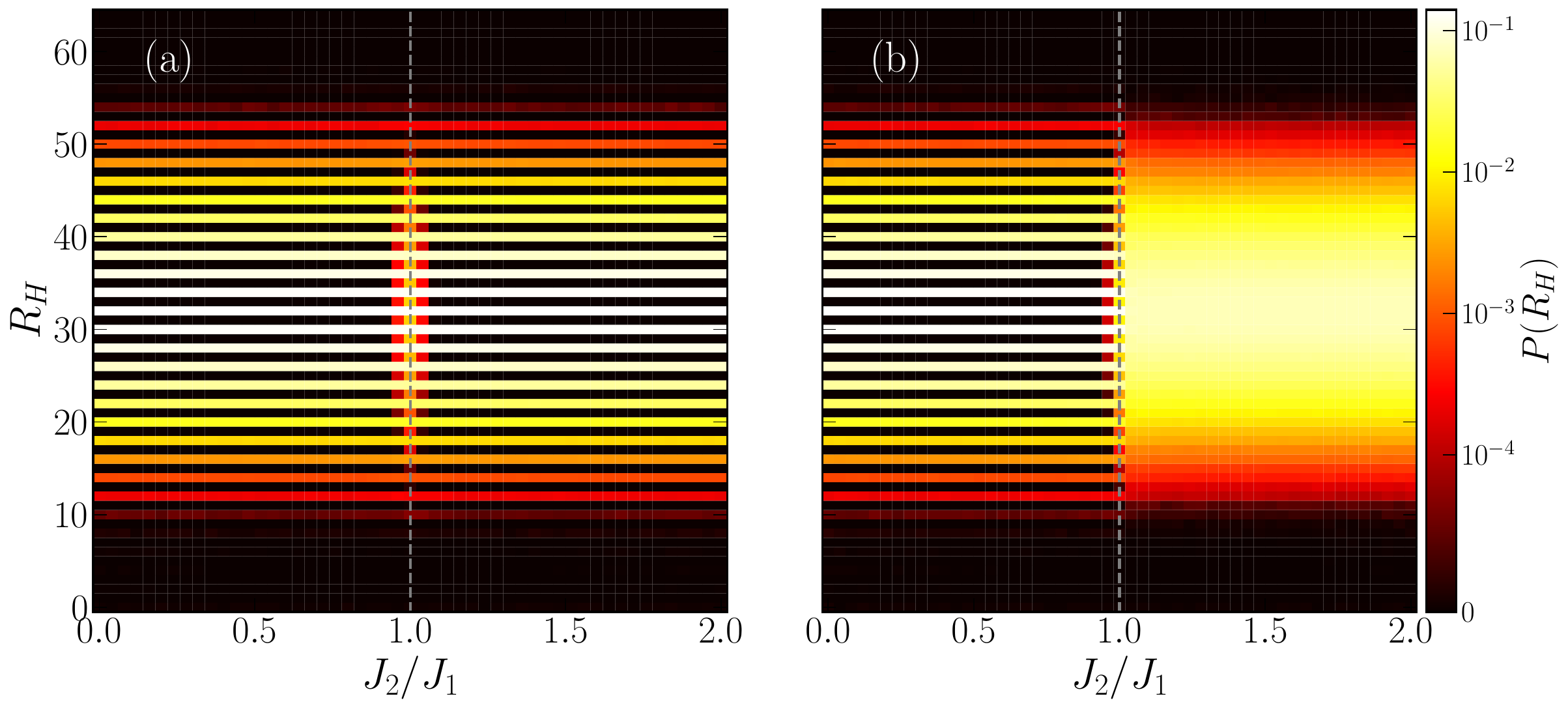}
    \caption{The evolution of probability distribution $P(r_{H})$ as a function of $J_2/J_1$ for SSH Heisenberg model in Eq.~\eqref{eq:ssh_heisenberg} with $L=128$. (a) For the PBC case, $P(r_{H})$ exhibits a ``stripy'' pattern in both gapped phases ($J_2/J_1 \neq 1$). (b) For the OBC case, the ``stripy'' pattern is prominent only in the topologically trivial phase ($J_2/J_1 < 1$) and abruptly vanishes in the non-trivial phase ($J_2/J_1 \ge 1$). 
    }
    \label{fig:ssh_pr_heatmap}
\end{figure}

Since $\mathrm{Var}(r_H)$ is fundamentally tied to local real-space observables, it is expected to be insensitive to symmetry-protected topological (SPT) phases and their transitions~\footnote{The SPT phase transition of the SSH model is accompanied by a bulk gap closing. Therefore, one can still see the signal of an anomaly at the transition point by checking real-space correlations, which are power-law here and exponential elsewhere. However, the complete description, including symmetry-breaking picture and critical exponents, is missing.}. Nevertheless, we begin our numerical investigation by directly examining the central object of our framework: the full probability distribution of Hamming distances. The numerical results in Fig.~\ref{fig:ssh_pr_heatmap} reveal that $P(r_H)$ for the SSH model exhibits a stark contrast with the TFIM [Figs.~\ref{fig:Pr_1D}(a) and \ref{fig:Pr_2D}(b)] and a strong dependence on boundary conditions. As predicted, the variance $\mathrm{Var}(r_H)$—merely the width of this distribution—does not serve as a marker for the different SPT phases. However, signals of the two phases and their transition may still be extractable from the distribution itself. 

\begin{figure}[t!]
    \centering
    \includegraphics[width=\linewidth]{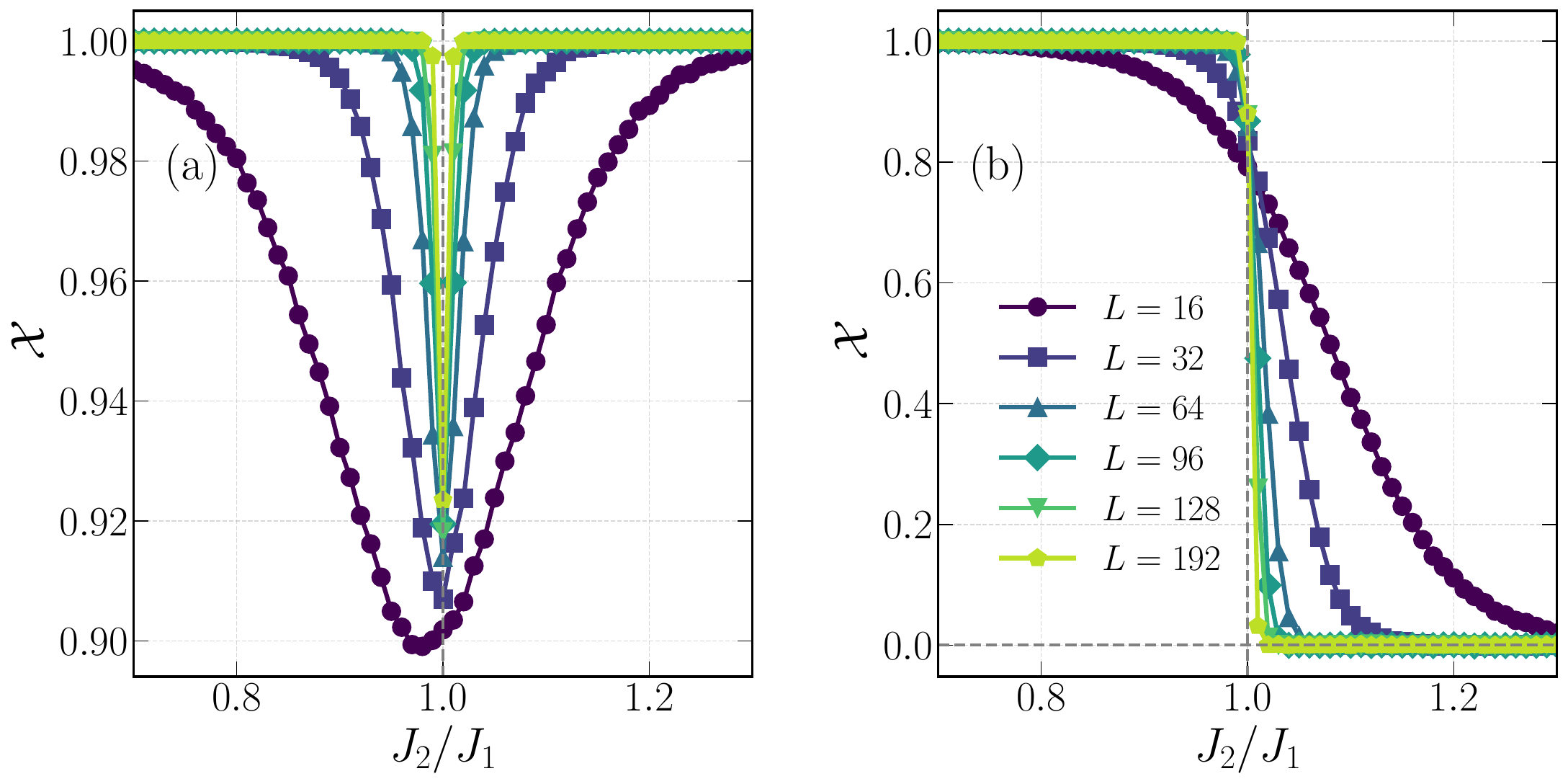}
    \caption{The parity index $\mathcal{X}$ as a function of the bond ratio $J_2/J_1$ for the alternating Heisenberg chain in Eq.~\eqref{eq:ssh_heisenberg}, with various system sizes $L$. The vertical dashed line marks the critical point at $J_2/J_1=1$. (a) The quantity $\mathcal{X}$ exhibits a sharp dip at this critical point with PBC. (b) For the OBC case, $\mathcal{X}$ serves as an excellent indicator for the topological phase transition, clearly distinguishing the trivial phase($\mathcal{X} \approx 1$) from the non-trivial phase ($\mathcal{X} \approx 0$). The two panels share the same legend.}
    \label{fig:ssh_parity_index}
\end{figure}

Specifically, a striking even-odd ``stripy'' pattern is observed in the geometric landscape of $P(r_H)$ (Fig.~\ref{fig:ssh_pr_heatmap}), which motivates a quantitative analysis of this characteristic, i.e., the parity index $\mathcal{X}$
\begin{equation}
    \mathcal{X} \equiv P(R_H=\text{even}) - P(R_H=\text{odd}),
    \label{eq:def_X}
\end{equation}
where $R_H = N r_H$ is the non-normalized Hamming distance between configurations. The numerical results for $\mathcal{X}$ as a function of the bond ratio $J_2/J_1$ are presented in Fig.~\ref{fig:ssh_parity_index} for both boundary conditions across a range of system sizes.

The distinct behavior of $\mathcal{X}$ has an intuitive interpretation in configuration space. In the disordered state, where configurations with different local structures are statistically uncorrelated, the distance between any two configurations is equally probable to be even or odd. This picture can be extended to a gapless ground state, minimal thermal fluctuations - as our SSE simulations are performed at a low but nonzero $T=1/L$ - produce a continuous probability distribution $P(r_H)$ devoid of even-odd modulation. In contrast, for the gapped dimerized ground state, the dominant-weight configurations correspond to dimers of spin singlets. The difference between any pair of such configurations involves an even number of differing single-site spins. Consequently, the statistics of even distances dominate $P(r_H)$, resulting in an even-odd stripy structure and parity index $\mathcal{X}\approx1$. 

The results of the SSH Heisenberg model under PBC nicely match the above intuitive explanation, with $\mathcal{X}<1$ at the gapless point ($J_2/J_1=1$) and $\mathcal{X}=1$ in both dimerized phases, as shown in Fig.~\ref{fig:ssh_parity_index}(a). For the OBC case, however, the parity index $\mathcal{X}$ acts as a remarkably clear non-local order parameter for the SPT phase transition, as shown in Fig.~\ref{fig:ssh_parity_index}(b). Deep in the trivial phase ($J_2/J_1 < 1$), where the chain is fully dimerized into singlet pairs, $\mathcal{X}$ approaches 1. As the system is tuned across the critical point into the non-trivial phase ($J_2/J_1 > 1$), $\mathcal{X}$ drops sharply toward zero. 

To elucidate this behavior, we establish an analytical connection between the parity index and a non-local, string-like physical operator. Considering the sum over all pairs of independently sampled configurations, $|s\rangle$ and $|s'\rangle$, we can write:
\begin{equation}
    \mathcal{X} = \sum_{s, s'} P(s) P(s') (-1)^{R_H(s, s')}.
\end{equation}
The term $(-1)^{R_H(s, s')}$ can be further expressed in terms of the spin values $s_i, s'_i \in \{+1, -1\}$ themselves. Since flipping a spin changes its sign, the product of all spins in configuration $|s'\rangle$ is related to that of $|s\rangle$ by $\prod_i s'_i = (\prod_i s_i) \cdot (-1)^{R_H(s, s')}$. This gives the identity $(-1)^{R_H(s, s')} = (\prod_i s_i)(\prod_i s'_i)$. Substituting this into the expression for $\mathcal{X}$ allows the sum to be factorized:
\begin{align}
    \mathcal{X} &= \sum_{s, s'} P(s) P(s') \left( \prod_i s_i \right) \left( \prod_i s'_i \right) \nonumber \\
    &= \left( \sum_s P(s) \prod_i s_i \right) \left( \sum_{s'} P(s') \prod_i s'_i \right) \nonumber \\
    &= \left\langle \prod_i \hat{\sigma}_i^z \right\rangle^2 = \left\langle \hat{P}_z\right\rangle^2.
    \label{eq:X_identity}
\end{align}
Thus, the distance-parity imbalance $\mathcal{X}$ is precisely the squared expectation value of the global $Z_2$ parity operator, $\hat{P}_z = \prod_i \hat{\sigma}_i^z$.

For the SSH Heisenberg model in Eq.~\eqref{eq:ssh_heisenberg}, the parity operator $\hat{P}_z$ commutes with the Hamiltonian with eigenvalues $p_z=\pm 1$. It follows that any non-degenerate eigenstate of the system exhibits $\mathcal{X} = \left\langle \hat{P}_z\right\rangle^2 = 1$. This explains why the fully dimerized ground state has $\mathcal{X}=1$ in two cases: all $J_2/J_1\neq 1$ under PBC and the $J_2/J_1<1$ regime under OBC. At the critical point $J_2/J_1=1$ under PBC, the gapless groundstate is blurred under minimal thermal fluctuations and one has $\mathcal{X}$ away from $1$. In contrast, the topologically non-trivial phase ($J_2/J_1>1$) under OBC features two unpaired, dangling spins at the chain ends. The resulting vanishing parity index $\mathcal{X}=0$ can be understood from two complementary perspectives: (i) the free end spins directly force the expectation value $\hat{P}_z$ to zero, and (ii) the ground state becomes 4-fold degenerate. Thus, the parity index $\mathcal{X}$, a quantity extracted directly from the geometry of the distance distribution $P(r_H)$, serves as a powerful non-local probe for detecting the SPT phase transition.

\section{Conclusion and Discussion}
\label{sec:discussion}

This work develops a geometric framework for characterizing phase transitions by analyzing the collective statistical properties of microscopic configurations. At its core, we shift the perspective from traditional real-space observables to the geometry of high-dimensional configuration space, demonstrating how universal critical behavior is encoded in the distance distribution $P(r_H)$ between sampled configurations.

We establish a rigorous analytical connection between the geometry of configuration-space statistics and real-space critical physics. Specifically, we derived that the standard deviation $\sqrt{\mathrm{Var}(r_H)}$ of Hamming distances follows a universal scaling law $\sim L^{-2\beta/\nu}$, which emerges from an exact mapping between Hamming distance statistics and real-space correlation functions. Theoretically, this scaling law holds when two conditions are met: the system must exhibit zero magnetization in the sampled basis and satisfy the dimensionality/criticality criterion $4\beta/\nu <d$. Our analytical framework provides a comprehensive explanation for the previous numerical findings in various classical spin systems~\cite{Su2023,Su2024} and is further validated in this work through the study of the quantum phase transition of the TFIM. The analytical connection between the configuration-space and real-space probes would also provide helpful physical insight into the ML approaches tackling phase transitions in configuration space~\cite{Rodriguez-Nieva2019,santos1,Santos2021prxq} and the Fock space landscape of the MBL~\cite{Logan2019,Sutradhar2022,Guo2021,Yao2023,Roy2024}.

We also explore the capacity of configuration-space measurements to transcend the limitations of real-space local observables. First, we introduce an information-geometric perspective based on the Fisher information metric for the $P(r_H; h)$ manifold. This approach complements the basis-dependent scaling analysis and reveals a key advantage: the Fisher information exhibits a basis-independent, critical divergence at phase transitions, unlike conventional local order parameters. We numerically demonstrate this universality for the TFIM across different bases, highlighting its potential value for probing transitions where traditional order parameters are ambiguous or undefined. Furthermore, we extend our analysis to the SSH Heisenberg model, which hosts symmetry-protected topological (SPT) physics beyond the LGW paradigm. By introducing a parity index derived from $P(r_H)$, we show that the configuration-space geometry successfully characterizes both the SPT phase and its transition.

Looking forward, this work establishes a foundational link between critical phenomena and configuration-space geometry—an insight demanding extensive theoretical and numerical exploration. Crucially, the second-order moment studied here, i.e., the standard deviation of $P(r_H)$, encodes physics equivalent to a real-space two-point correlator, functioning as a quasi-local probe of symmetry-breaking transitions. This invites a natural question: Could higher-order geometric invariants, characterizing multi-configuration distances or topological features of $P(r_H)$, access non-local correlations and describe phase transitions beyond LGW paradigms? The very preliminary exploration in Sec.~\ref {sec:nonlocal_probes} has demonstrated its potential in revealing signatures of unconventional criticality inaccessible to local measurements. We expect the present work to inspire future works along this line, especially in characterizing novel phases such as interacting SPT phases and quantum spin liquid. Specifically, considering the spin liquid formed by the randomly paired dimers in real space, the parity index defined in Sec.~\ref{sec:NL_parity_index} functions as a powerful probe of tackling phase transitions between complex magnetic order and quantum spin liquid phase. 
 Parallelly, developing efficient computational strategies is essential to harness these geometric constructs as novel probes of criticality. 

\section*{Acknowledgments}
We thank Y.-C. Wang for helpful discussion on the criticality of spin models. This research was supported by the National Natural Science Foundation of China (grant nos.. 12174167, 12474139, 12247101), the Fundamental Research Funds for Central Universities (Grant No. lzujbky-2024-jdzx06), and the Natural Science Foundation of Gansu Province (No. 22JR5RA389).

\appendix

\section{Numerical details}
\label{app:num_details}

This appendix provides technical details of our SSE QMC simulations and the procedure for calculating the Hamming distance statistics.

\subsection{SSE-QMC algorithm}

Our simulations are based on the SSE-QMC method, a powerful and widely used numerical technique for quantum spin systems with no sign problem. 
Besides, to ensure sampling validity and efficiency for both Hamiltonians studied in this work, we employed optimized update strategies:
\begin{itemize}
    \item For $\hat{\sigma}^z$-basis measurements, we simulate the original Hamiltonian $H_\mathrm{TFIM}$ in Eq.~\eqref{eq:H_TFIM} with an efficient \textit{quantum cluster algorithm}~\cite{Sandvik_SSE_TFIM}. This algorithm is highly effective at updating configurations involving off-diagonal transverse-field terms and mitigating critical slowing down.
    \item For $\hat{\sigma}^x$-basis measurements, we simulate the rotated Hamiltonian $H_{HB}$ Eq.~\eqref{eq:H_rotated} with a powerful and general \textit{directed loop algorithm}~\cite{Sandvik_SSE_DL}. This algorithm provides a versatile mechanism for updating configurations by constructing and flipping "loops" of spin histories, which is well-suited for systems with off-diagonal spin couplings.
\end{itemize}

For each set of model parameters (system length $L$ and transverse field $h$), we perform a series of independent QMC simulations starting from different initial configurations, i.e., the so-called \textit{bins}, to facilitate robust error analysis. A typical simulation run for a single bin consists of:
\begin{itemize}
    \item Thermalization: For one-dimensional systems and smaller two-dimensional systems, we first perform $2 \times 10^4$ QMC sweeps to ensure the system reaches thermal equilibrium. For larger 2D systems (e.g., $L\geq64$), we perform $10^4$ thermalization sweeps.
    \item Measurement: Following thermalization, for 1D and smaller 2D systems, we perform an additional $4 \times 10^4$ sweeps for measurements. For larger 2D systems, we perform $10^4$ measurement sweeps.
\end{itemize}
For the results presented in this work, we typically use $\mathcal{N}_b = 64$ independent bins to ensure that the statistical errors, calculated as the standard error of the mean over these bins, are reliable and small.

\subsection{Configuration Sampling and Distance Calculation}

The core of our method relies on analyzing the statistics of distances between configurations sampled from the equilibrium ensemble. The procedure is as follows:

\begin{enumerate}
    \item Configuration extraction: During the measurement steps of each QMC bin, we extract snapshots of the system's configuration. For the SSE simulation method, there are $N_{l}$ \textit{imaginary-time slices} with $N_{l}\sim L^{z}$ in each measurement step. In this work, we randomly adopt one single spin configuration from $1\sim N_{l}$ \textit{imaginary-time slices} of the propagated state, to approximate the full distribution $P(s)$. To ensure the configurations are statistically independent, we perform 20 QMC update sweeps between each extraction. For each bin, we typically collect $\mathcal{N}_c = 2000$ such independent configurations.

    \item Building the configuration pool: We pool all configurations collected from all $\mathcal{N}_b$ bins, resulting in a total set of $\mathcal{N}_s = \mathcal{N}_b \times \mathcal{N}_c$ (e.g., $64 \times 2000 = 128,000$) configurations. This set, denoted as $\{s\}$, represents a faithful statistical sample of the system's equilibrium state.

    \item Distance calculation: From the configuration pool $\{s\}$, we generate a dataset of distances $\{r_H\}$. We employ a random sampling approach to construct this dataset, rather than calculating all possible pairs. Specifically, we randomly draw $10 \times \mathcal{N}_s$ pairs of configurations from the pool and compute their normalized Hamming distance $r_H$. This approach has three main advantages: i) strikes a good balance between computational efficiency and statistical accuracy, as the final precision is primarily limited by the number of independent configurations $\mathcal{N}_s$; ii) it saves significant storage space; and iii) random pairing effectively reduces potential autocorrelation effects from pairs drawn from within the same bin.

    \item Statistical analysis: With the generated dataset of distances $\{r_H\}$, we can accurately compute its probability distribution $P(r_H)$ by simple histogramming. The mean (first moment) $\langle r_H \rangle$ and standard deviation (second moment) $\sqrt{\mathrm{Var}(r_H)}$ of the distribution are directly computed from the dataset $\{r_H\}$.
\end{enumerate}
This systematic and robust procedure ensures that the calculated Hamming distance statistics are a reliable and accurate representation of the system's properties in the configuration space.

\begin{figure}[!t]
    \centering
    \includegraphics[width=0.95\linewidth]{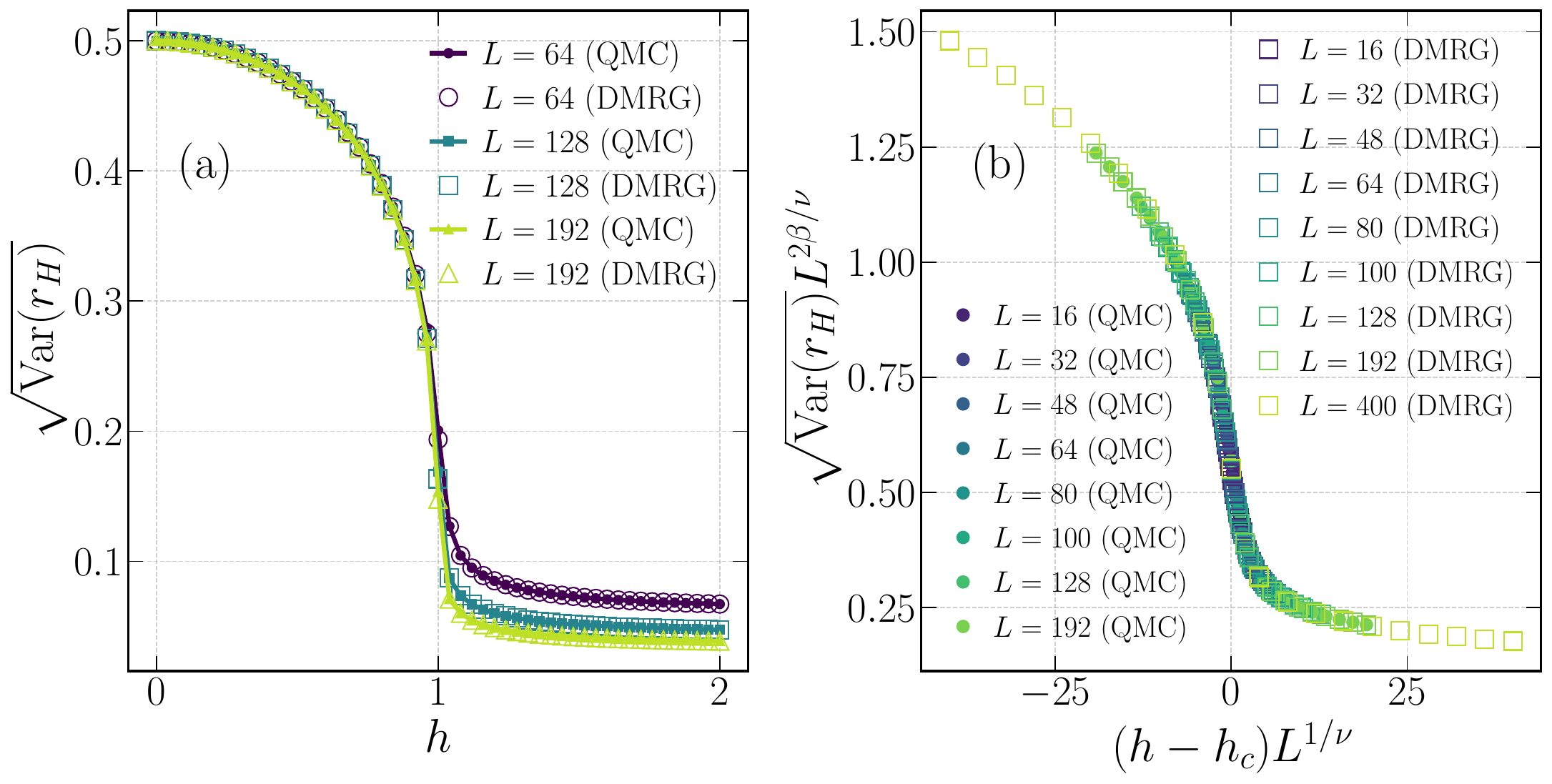} 
    \caption{
        DMRG benchmark and scaling validation for the 1D TFIM. (a) Cross-validation of $\sqrt{\mathrm{Var}(r_{H})}$. QMC data (solid markers and lines) are compared against DMRG calculations (larger empty markers) for several system sizes. (b) Data collapse of both QMC (small solid markers) and DMRG (larger empty markers) data, scaled using the theoretical exponents of the 2D Ising universality class.
    }
    \label{fig:dmrg_benchmark}
\end{figure}

\subsection{DMRG benchmark and data collapse validation}

To rigorously benchmark our findings, we compare our numerical results from SSE QMC simulations against high-precision DMRG calculations for the 1D TFIM. This validation, presented in Fig.~\ref{fig:dmrg_benchmark}, proceeds in two steps.

First, we perform a cross-validation of the observable $\sqrt{\mathrm{Var}(r_{H,z})}$. As shown in Fig.~\ref{fig:dmrg_benchmark}(a), the QMC data (solid markers with lines) for various system sizes show excellent agreement with the DMRG results (hollow markers), confirming the accuracy of our QMC protocol in capturing the ground-state physics. Here, the curves of the DMRG data are reconstructed from real-space correlations according to Eq.~\eqref{eq:var_rH}, further verifying the analytical derivations in Sec.~\ref{sec:derivations}. The minor deviations near the critical point are expected to be finite-temperature effects in the QMC simulation.

Second, we validate the universal scaling law itself. Figure~\ref{fig:dmrg_benchmark}(b) shows the data collapse for both QMC (small solid markers) and DMRG (large hollow markers) data from the critical region. By scaling the axes with the known theoretical exponents of the 2D Ising universality class ($h_c=1.0$, $\nu=1.0$, and $2\beta/\nu=0.25$), both datasets collapse onto a single universal curve. The excellent collapse of two distinct numerical methods with the theoretical prediction provides a powerful confirmation of our proposed scaling law.

\section{Scaling Procedure and Data Collapse}
\label{app:scaling_details}

The critical point $h_c$ and the critical exponent ratios $2\beta/\nu$ and $1/\nu$ are numerically determined by optimizing the data collapse of the standard deviation $\sqrt{\mathrm{Var}(r_{H})}$. This procedure is based on the finite-size scaling ansatz near a continuous phase transition in Eq.~\eqref{eq:scaling_con}. The goal is to find the set of parameters $\{h_c, 2\beta/\nu, 1/\nu\}$ that makes all numerical data points collapse onto a single smooth curve when plotted as $Y = \sqrt{\mathrm{Var}(r_{H})} L^{2\beta/\nu}$ versus $X = (h-h_c)L^{1/\nu}$.

To quantify the quality of the data collapse for a given set of trial parameters, we use the cost function defined as~\cite{Jan2020,Aramthottil2021,liang2023disorder,Su2023,Su2024}: 
\begin{align}
    C_{\mathrm{scaling}}=\frac{\sum_j |Y_{j+1}-Y_j|}{\max\{Y_j\}-\min\{Y_j\}}-1.
    \label{eq:cost_function_app}
\end{align}
Here, the set $\{ (X_j, Y_j) \}$ represents all rescaled data points from all system sizes $L$. By sorting these points such that the $X_j$ values are in non-decreasing order ($X_j \leq X_{j+1}$), the cost function $C_{\mathrm{scaling}}$ measures the ``smoothness'' of the resulting curve. A smaller value of $C_{\mathrm{scaling}}$ corresponds to better data collapse. In the present work, we aim for the optimal critical parameters that yield the minimum value of $C_{\mathrm{scaling}}$.

\section{Details of TFIM in the Transverse Basis}
\label{app:x_basis_analysis}

This appendix analyzes the key physical observables in the transverse ($\hat{\sigma}^x$) basis for the 1D TFIM. Although the observables are discussed in the $\hat{\sigma}^x$ basis, the corresponding SSE simulations are performed using the hard-core boson representation of the Hamiltonian in Eq.~\eqref{eq:H_rotated}. For our analysis, these representations are equivalent, and we do not distinguish between them.

\begin{figure}[!t]
    \centering
    \includegraphics[width=\linewidth]{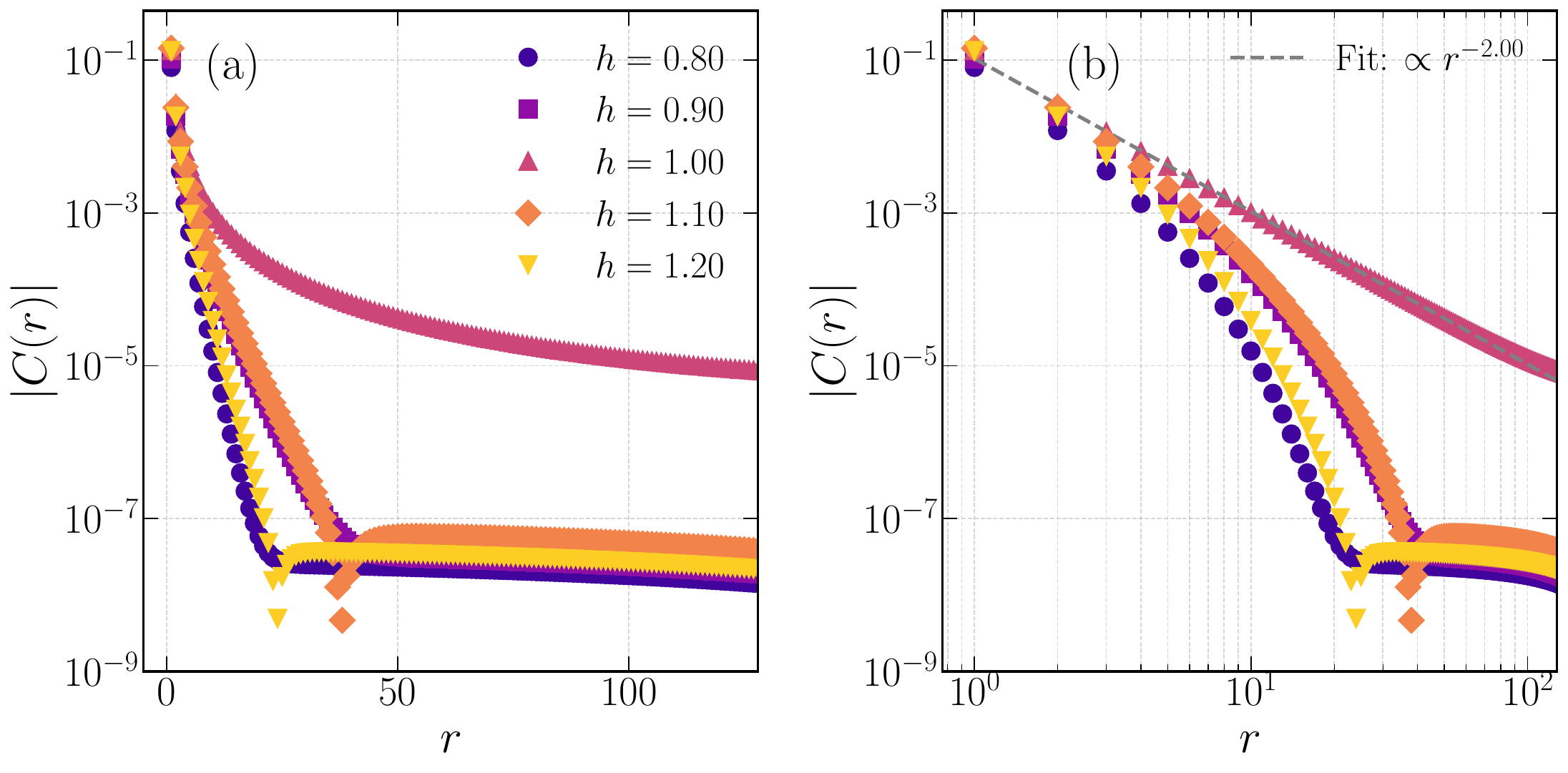}
    \caption{Decay of the correlation function $|C(r)|$ in the $\sigma_x$ basis for the 1D TFIM. (a) On a semi-log plot, the data for non-critical fields ($h \neq 1.0$) exhibit linear trends, characteristic of exponential decay. (b) On a log-log plot, the data at the critical point falls on a straight line, confirming its power-law nature with a fitted exponent $p \approx 2.00$. Here the data are obtained by DMRG for $L=400$.}
    \label{fig:appendix_x_basis_decay}
\end{figure}

The correlation function of the TFIM along the $\hat{\sigma}^x$ direction is well established theoretically. Away from the critical point ($h \neq h_c$), correlations decay exponentially. Precisely at the quantum critical point ($h = h_c$), the correlations exhibit power-law decay with a large exponent, $C(r) \sim r^{-2}$~\cite{Pfeuty1970}. This distinct behavior is verified using high-precision DMRG calculations for a long chain with $L=400$, as displayed in Fig.~\ref{fig:appendix_x_basis_decay}. 

\begin{figure}[b!]
    \centering
    \includegraphics[width=\linewidth]{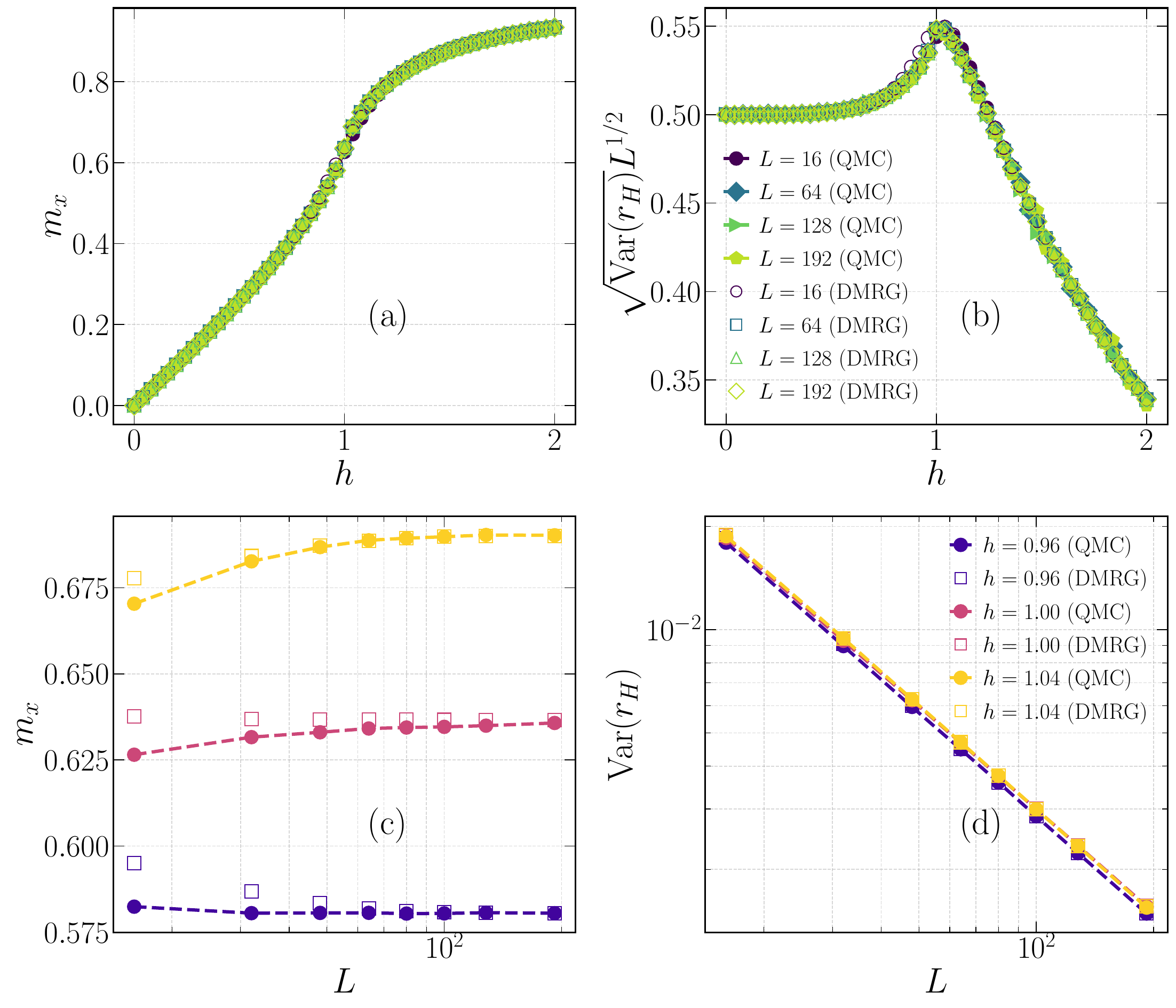} 
    \caption{Comprehensive analysis of observables in the $\hat{\sigma}^x$ basis. (a) Transverse magnetization $m_x$. (b) Non-critical data collapse of $\sqrt{\mathrm{Var}(r_{H})} L^{1/2}$. (c) Weak finite-size dependence of $m_x$. (d) Trivial scaling of $\mathrm{Var}(r_{H}) \sim L^{-1}$. Numerical results from QMC and DMRG are both preseted for comparison.}
    \label{fig:app_x_basis_ observables}
\end{figure}

As shown in Fig.~\ref{fig:app_x_basis_ observables}(a) and (c), the magnetization $m_x$ exhibits non-critical character: its value evolves smoothly with the tuning parameter $h$ and displays only weak finite-size effects, converging rapidly to the thermodynamic limit. This behavior is characteristic of an observable that does not serve as an order parameter for the transition. 


Given the findings of rapid power-law decay of $C(r)^{-2}$ and the noncritical $m_x$, we can now elucidate the scaling behavior of the distance variance $\mathrm{Var}(r_H)$. At the critical point, all three terms in Eq.~\eqref{eq:var_rH} are found to share the same trivial scaling. The first term scales as $L^{-d}$ by definition, and the rapid $C(r) \sim r^{-2}$ decay ensures that the sums in the second and third terms also converge to $L$-independent constants, leading to an $L^{-d}$ scaling for them as well. This results in a total leading power of $\mathrm{Var}(r_H) \sim L^{-d} = L^{-1}$. This scaling form also holds for the regime away from the critical point, where the correlation decays exponentially. In this way, the second moment of $P(r_H)$ loses criticality along the $\hat{\sigma}^x$ direction.

Our numerical results perfectly validate this theoretical prediction. Figure~\ref{fig:app_x_basis_ observables}(d) shows that $\mathrm{Var}(r_H)$ indeed follows a power law with a slope near -1. Furthermore, Fig.~\ref{fig:app_x_basis_ observables}(b) shows that the rescaled standard deviation, $\sqrt{\mathrm{Var}(r_H)} L^{1/2}$, collapses excellently for all system sizes. We also make a direct comparison of QMC and DMRG results: although QMC result in the inset of Fig.~\ref{fig:x_basis_analysis}(a) has less accuracy and drops rapidly to an error background, it is sufficient to catch the correct scaling behavior of $\sqrt{\mathrm{Var}(r_H)}$.

\section{Fisher Information of SSH Heisenberg model and SPT phase transition}
\label{sec:ssh_fi_analysis}

\begin{figure}[t!]
    \centering
    \includegraphics[width=\linewidth]{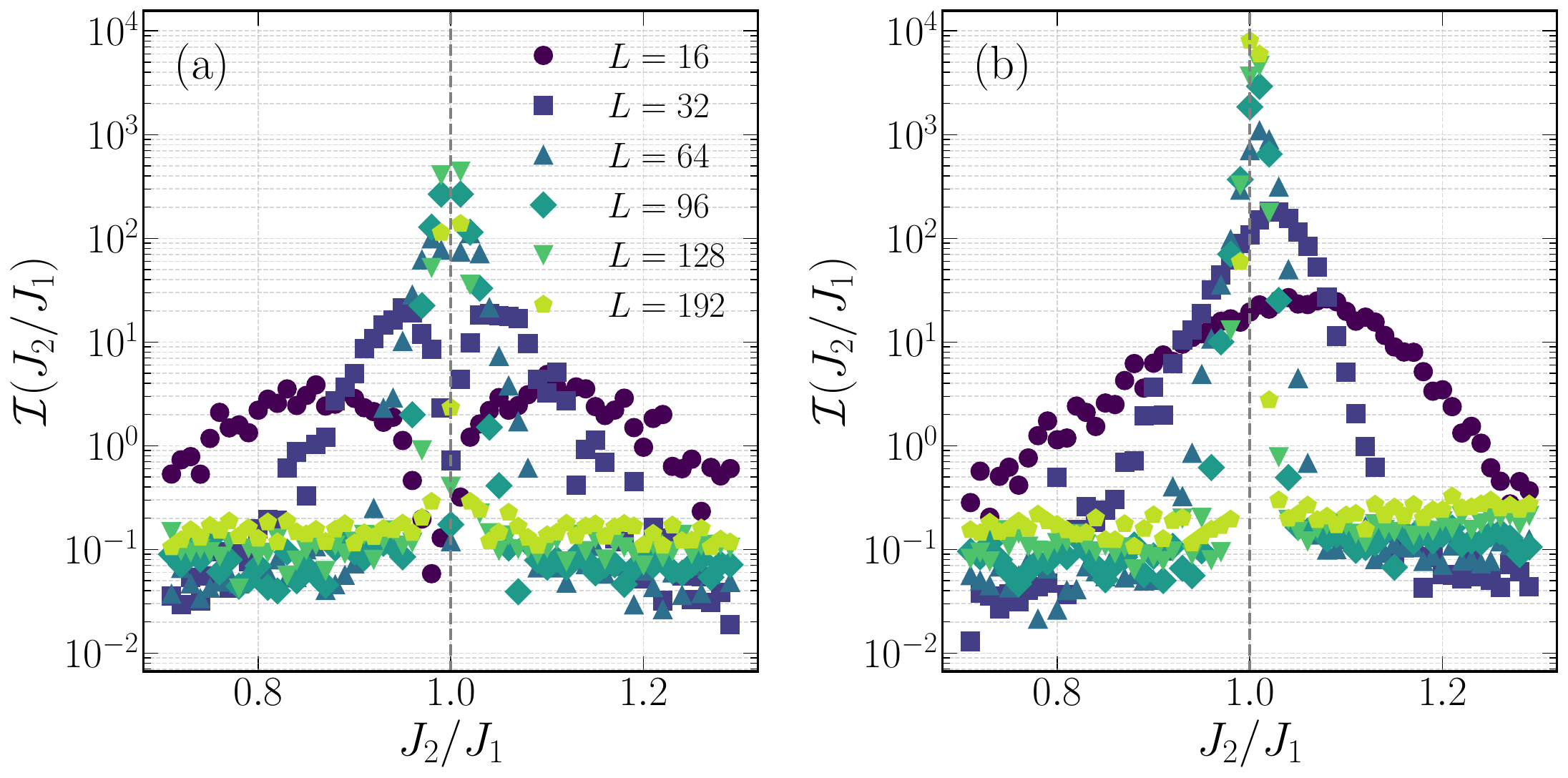}
    \caption{Detection of the SPT transition using Fisher Information, $\mathcal{I}(J_2/J_1)$, for various system sizes $L$. The vertical dashed line at $J_2/J_1=1$ marks the known critical point.
    (a) Under periodic boundary conditions (PBC), the FI displays a characteristic "double-peak-and-valley" structure, with a distinct minimum precisely at the critical point.
    (b) Under open boundary conditions (OBC), the FI exhibits a single, sharp peak that grows and narrows with increasing system size, unambiguously identifying the critical point.}
    \label{fig:ssh_fisher_info}
\end{figure}

As revealed by the heatmaps in Fig.~\ref{fig:ssh_pr_heatmap}, the geometric structure of $P(r_H)$ undergoes a significant alteration precisely at the SPT phase transition point $J_2/J_1=1$. It follows naturally that the Fisher information, introduced in Sec.~\ref{sec:NL_fisher_information} and defined in Eq.~\eqref{eq:FI} as a measure of this geometric change, would serve as a sensitive indicator of the transition. Our numerical results for the SSH Heisenberg model, presented in Fig.~\ref{fig:ssh_fisher_info}, confirm this: the Fisher information provides a clear and accurate signature of the phase transition under both PBC and OBC. However, the qualitative features of the signal are strikingly dependent on the boundary conditions.

In the case of OBC, as shown in Fig.~\ref{fig:ssh_fisher_info}(b), the Fisher information exhibits the standard hallmarks of a quantum phase transition. A single sharp peak emerges precisely at the critical point. As the system size $L$ increases, this peak grows in height and narrows, signaling a divergence in the thermodynamic limit. This behavior originates directly from the geometric transformation of $P(r_H)$ observed in Fig.~\ref{fig:ssh_pr_heatmap}: at criticality, the distribution undergoes a singular change from a parity-imbalanced stripey structure to a parity-symmetric one. The Fisher information serves as a sensitive metric that precisely pinpoints this moment of maximal geometric change.

In the case of PBC, as shown in Fig.~\ref{fig:ssh_fisher_info}(a), the Fisher information reveals a more complex, bimodal signature characterized by a central valley flanked by two peaks. A distinct minimum is located precisely at the critical point, with the flanking peaks growing sharper and taller as the system size increases. This non-trivial structure directly reflects the constrained evolution of $P(r_H)$ under PBC. As established earlier, the critical point corresponds to a gapless state of mixed character. The valley at $J_2/J_1=1$ signifies that this state represents a point of relative structural stability, where the geometry of $P(r)$ is least sensitive to variations in the tuning parameter. Conversely, the flanking peaks mark the regions of maximal geometric change, corresponding to the transitions from the fully dimerized gapped phases into the critical regime. Despite its qualitatively different shape from that in the OBC case, this unique bimodal signature provides an unambiguous and precise marker for the critical point.

In conclusion, the Fisher Information provides independent validation of the SPT transition location identified by the parity index $\mathcal{X}$ in Sec.~\ref{sec:NL_parity_index}. The emergence of qualitatively distinct signatures for OBC and PBC is not a contradiction, but rather a profound demonstration of its sensitivity to the system's global topology. This firmly establishes it as a robust and versatile component of our framework, capable of detecting phase transitions that lie outside the conventional paradigm of spontaneous symmetry breaking.

\bibliography{draft_v2}

\end{document}